\documentclass[11pt,a4paper]{article}
\usepackage{epsfig,axodraw}
\usepackage{array,cite,amsmath,amssymb}
\usepackage[flushleft]{threeparttable}
\usepackage{subcaption}
\usepackage{kantlipsum}
\usepackage{float}
\usepackage[usenames,dvipsnames]{xcolor}
\usepackage[breakable, theorems, skins]{tcolorbox}
\tcbset{enhanced}
\usepackage{booktabs}
\usepackage{caption}
\usepackage{threeparttable}
\usepackage{multirow}
\usepackage{listings}
\definecolor{DarkMagenta}{rgb}{0.55,0.0,0.55}
\newcommand{\newc}{\newcommand}
\newc{\gev}{\,GeV}
\newcolumntype{M}[1]{>{\centering\arraybackslash}m{#1}}
\newcolumntype{N}{@{}m{0pt}@{}}
\newc{\mev}{\,MeV}
\newc{\ra}{\rightarrow}
\newc{\rpv}{$\mathrm{\not\!R_p}$}
\newc{\rp}{$\mathrm{R_p}$}
\newc{\real}{\mathcal{R}e}
\newc{\alsm}{{\displaystyle \sum_{\alpha=1,2}}}
\newc{\besm}{{\displaystyle \sum_{\beta=1,2}}}
\newc{\al}{\alpha}
\newc{\sgn}{\mr{sgn}\,}
\newc{\be}{\beta}
\newc{\ga}{\gamma}
\newc{\de}{\delta}
\newc{\sla}{\!\!\!\!\!\not\:\:\!}
\newc{\slab}{\!\!\!\!\!\not\,\,\,}
\newc{\slac}{\!\!\!\!\!\!\!\not\,\,\,\,}
\newc{\met}{$\not\!\!E_T$}
\newc{\cw}{\cos\theta_W}
\newc{\sw}{\sin\theta_W}
\newc{\ssw}{\sin^2\theta_W}
\newc{\ccw}{\cos^2\theta_W}
\newc{\cbe}{\cos\beta}
\newc{\sbe}{\sin\beta}
\newc{\ort}{\frac1{\sqrt{2}}}
\newc{\sh}{\hat{s}}
\newc{\uh}{\hat{u}}
\newc{\tha}{\hat{t}}
\newc{\sa}{\sin\al}
\newc{\ca}{\cos\al}
\newc{\mz}{M_{\mr{Z}}}
\newc{\mw}{M_{\mr{W}}}
\newc{\bv}{$\mathrm{\not\!B}$}
\newc{\lv}{$\mathrm{\not\!L}$}
\newc{\beq}{\begin{equation}}
\newc{\eeq}{\end{equation}}
\newc{\ie}{{\it i.e.\/}\ }
\newc{\lam}{\lambda}
\newc{\cht}{\tilde{\chi}}
\newc{\glt}{\tilde{g}}
\newc{\upt}{\tilde{u}}
\newc{\qkt}{\tilde{q}}
\newc{\elt}{\tilde{\ell}}
\newc{\hgt}{\tilde{H}}
\newc{\nut}{\tilde{\nu}}
\newc{\dnt}{\tilde{d}}
\newc{\ftl}{\mr{\tilde{f}}}
\newc{\psb}{\bar{\psi}}
\newc{\rtt}{2^{1/2}}
\newc{\mut}{\tilde{\mu}}
\newc{\mr}{\mathrm}
\newc{\bath}{\bar{\theta}}
\newc{\tht}{\theta}
\newc{\JC}{{\bf J}}
\newc{\lra}{\longrightarrow}
\newc{\eg}{{\it e.g.\  }}
\newc{\barr}{\begin{eqnarray}}
\newc{\earr}{\end{eqnarray}}
\newc{\me}{\mathcal{M}}
\newc{\dbm}{\partial_\mu}
\newc{\dbmu}{\stackrel{\leftrightarrow\  }{\partial^\mu}}
\newc{\sgm}{\sigma_\mu}
\newc{\captionB}[2]{\caption[{#1}]{{\small {#2}}}}
\newc{\ahref}[2]{#2}
\pdfoutput=1 

\usepackage{array,cite,amsmath,color}
\usepackage{xspace}
\usepackage{bbm,subcaption}
\usepackage{slashed}

\usepackage{jheppub} 
\title{Coherent and Dissociative Photon-Induced Production in Herwig 7}

\author[a]{L.~Grazette,}
\author[b,c]{K.~Lohwasser,}
\author[d]{M.R.~Masouminia}

\affiliation[a]{Department of Physics, University of Warwick, Coventry, UK}
\affiliation[b]{Department of Physics and Astronomy, University of Sheffield, Sheffield, UK}
\affiliation[c]{Theoretical Physics Department, CERN, Geneva, Switzerland}
\affiliation[d]{Institute for Particle Physics Phenomenology, Durham University, Durham, UK}

\emailAdd{L.Grazette@warwick.ac.uk}
\emailAdd{kristin.lohwasser@cern.ch}
\emailAdd{mohammad.r.masouminia@durham.ac.uk}

\abstract{ \small
We present a particle-level study of photon-induced production in \textsf{Herwig~7}, focusing on coherent, single-dissociative and double-dissociative proton configurations. The simulation combines the Budnev elastic photon flux with an inelastic photon PDF and uses internal \textsf{Herwig~7} matrix elements for $\gamma\gamma\to \ell^+\ell^-$ and $\gamma\gamma\to W^+W^-\to \ell^+\ell^-\nu\bar{\nu}$ as the central setup. Using a dedicated \textsf{Rivet} analysis, we compare hard-system kinematics, charged-particle activity, rapidity-gap observables and forward-remnant-sensitive distributions. We find that shower-setting changes generally have modest effects on hard-system distributions in well-populated regions, while coherent and dissociative samples differ most clearly in additional charged activity and rapidity-gap observables. These results provide a useful baseline for modelling semi-exclusive photon-induced measurements at the LHC and identify the observables most sensitive to dissociative activity in future generator comparisons.
}

\begin{document}
\noindent{\hfill \small MCnet-26-33}

\noindent{\hfill \small IPPP/26/73\\[0.1in]}
\maketitle
\flushbottom

\section{Introduction}
\label{sec:intro}

Photon-induced processes provide an important class of colour-singlet reactions at the LHC, particularly in measurements employing exclusive or semi-exclusive event selections. In such analyses, the signal is not identified only through the hard final state, for example a lepton pair or a diboson system, but also through the absence of additional hadronic activity, the presence of rapidity gaps, or the reconstruction of intact forward protons. Exclusive $\gamma\gamma\to \ell^+\ell^-$ production provides the cleanest benchmark channel, with measurements by CMS and ATLAS using event selections designed to suppress additional activity beyond the lepton pair~\cite{Chatrchyan:2011ci,Aad:2015bwa}. Photon-induced $W^+W^-$ production probes the same underlying $\gamma\gamma$ topology in a diboson final state and has been studied by CMS at 7 and 8 TeV, and observed by ATLAS at 13 TeV, using selections based on the leptonic $W$ decays and vetoes on additional charged particles~\cite{Chatrchyan:2013foa,CMS:2016exclWW,ATLAS:2020photonWW}. More recently, forward-proton measurements of high-mass lepton-pair production by CMS-TOTEM and ATLAS have demonstrated the importance of combining central-system kinematics with proton-tagging information~\cite{CMS-TOTEM:2018hki,ATLAS:2020wny}. The modelling of these observables is 
directly tied to the treatment of the incoming photons.

From the experimental point of view, the distinction between coherent, single-dissociative and double-dissociative photon-induced production is mainly a distinction in the forward and accompanying event activity, rather than in the central hard system. However, it crucially determines the cross-section that is experimentally accessible. Elastic, or coherent, photon emission is described in terms of the electromagnetic structure of the proton and leaves the emitting proton intact, whereas inelastic photon emission is associated with proton dissociation and is described through proton structure functions or photon PDFs~\cite{Budnev:1975poe,Drees:1988pp,Manohar:2016nzj,Manohar:2017eqh}. In coherent production, both protons remain intact, leading to the topology $pp\to p+X+p$, where $X$ denotes the central photon-induced final state. In single-dissociative production, only one proton remains intact, $pp\to p+X+Y$, while in double-dissociative production both protons dissociate, $pp\to Y_1+X+Y_2$. These configurations can therefore produce the same central final state, such as $\ell^+\ell^-$ or $W^+W^-$, but differ in the probability of observing forward protons, rapidity gaps, additional charged particles, and remnant activity in the detector~\cite{Harland-Lang:2016apc,Harland-Lang:2020veo}. Consequently, finite detector acceptance and exclusivity requirements can cause migrations between exclusive, semi-exclusive and inclusive event categories. In the ATLAS $\gamma\gamma\to W^+W^-$ fiducial selection, for example, single-dissociative production contributes 81\% of the cross-section prediction~\cite{ATLAS:2020photonWW}.

The single-dissociative component is particularly delicate because it combines features of both the coherent and fully dissociative limits. On one side of the event, an intact proton may be present, giving rise to a forward-proton tag or a rapidity-gap-like topology, while on the other side, the dissociating proton produces a remnant system whose particles may or may not enter the detector acceptance. This makes the classification of such events sensitive to remnant fragmentation, charged-particle thresholds, rapidity-gap definitions and soft-survival effects. As a result, single-dissociative events can easily migrate between experimentally exclusive, semi-exclusive and inclusive categories, even when the central photon-induced system is well reconstructed~\cite{Harland-Lang:2016apc,Harland-Lang:2020veo}.

Photon-induced events have been simulated using both dedicated tools and general-purpose event generators. Historically, calculations based on the equivalent-photon approximation, and in particular the \textsf{LPAIR} generator, provided the standard description of two-photon lepton-pair production in hadron collisions~\cite{Vermaseren:1982cz,Baranov:1991}. More recent specialised tools, such as \textsf{SuperChic} and \textsf{CepGen}, offer dedicated descriptions of central exclusive and photon-induced production, including elastic and inelastic photon emission, proton dissociation and, in the case of \textsf{SuperChic}, soft-survival effects relevant for rapidity-gap selections~\cite{Harland-Lang:2015cta,Harland-Lang:2020veo,Forthomme:2018sxa}. Photon-induced processes can also be studied in general-purpose frameworks such as \textsf{Pythia}~\cite{Sjostrand:2014zea,Bierlich:2022pfr,Helenius:2018uul}, \textsf{Sherpa}~\cite{Gleisberg:2008ta,Bothmann:2019yzt,Bothmann:2024yqu,Archibald:2008zza,Hoche:2023mkv}, \textsf{MadGraph} interfaced to a shower generator~\cite{Alwall:2014hca}, and \textsf{Herwig~7}~\cite{Bahr:2008pv,Bellm:2015jjp,Bellm:2019zci,Bewick:2023tfi,Bellm:2025pcw}. These tools offer complementary strengths: dedicated generators provide a more focused treatment of exclusive photon-induced kinematics and survival effects, while general-purpose generators allow the photon-induced hard process to be embedded in a full simulation of showering, beam remnants, hadronization and particle decays.

A remaining ambiguity in generator predictions is not only the normalisation of the photon-induced cross section, but also how different modelling choices affect the observables used to define exclusivity. The photon flux or photon PDF controls the rate and kinematics of the incoming photons, and hence the central hard system, while the treatment of proton dissociation controls the remnant system accompanying the hard process. In practice, however, these effects are not cleanly separated experimentally: charged-particle multiplicities, rapidity gaps and forward activity are sensitive to remnant fragmentation, showering, hadronization and soft-survival effects, as well as to detector acceptance and selection thresholds~\cite{Manohar:2016nzj,Manohar:2017eqh,Harland-Lang:2016apc,Harland-Lang:2020veo}. A useful generator study should therefore establish which distributions are mainly probing the photon-induced hard kinematics, and which are instead dominated by the modelling of dissociation and associated event activity.

In this paper we perform a controlled study of photon-induced production in \textsf{Herwig~7}. The aim is not simply to generate $\gamma\gamma$ hard processes, but to test how the coherent, single-dissociative and double-dissociative components are realised inside a full general-purpose event generator, including showering, beam-remnant treatment, hadronization and the resulting particle-level activity. This is the aspect for which \textsf{Herwig~7} is particularly useful: the same framework can be used to compare intact-proton photon emission, inelastic photon PDFs and the associated remnant fragmentation, while keeping the hard matrix element, analysis definition and non-photon modelling choices under direct control~\cite{Bahr:2008pv,Bellm:2015jjp,Bewick:2023tfi}.

We focus on two representative final states, $\gamma\gamma\to\ell^+\ell^-$ and $\gamma\gamma\to W^+W^-\to\ell^+\ell^-\nu\bar{\nu}$. The first provides a clean benchmark for the photon-induced kinematics, while the second probes a diboson topology  relevant to BSM searches. For each process we compare coherent, single-dissociative and double-dissociative configurations, and study observables sensitive to the hard photon-induced system, charged-particle activity, rapidity gaps and forward remnant structure. This allows us to identify which features are robust properties of the central $\gamma\gamma$ process and which are driven by the modelling of proton dissociation and the accompanying event activity.

The present study is therefore intended as a \textsf{Herwig~7} baseline and as a diagnostic of the modelling systematics internal to this framework. Comparisons to other generators, in particular \textsf{MadGraph} interfaced to \textsf{Pythia}, are important and natural follow-up studies, but are deliberately left outside the central scope of this paper. This separation allows us first to establish a reproducible Herwig-based reference before introducing differences from matrix-element generation, shower algorithms, hadronization models and alternative treatments of the underlying event.

The paper is organised as follows. In section~\ref{sec:pi-herwig}, we describe the definition of coherent, single-dissociative and double-dissociative photon-induced production in \textsf{Herwig~7}, including the photon-flux and photon-PDF choices, the internal matrix elements used as the baseline, and the treatment of showering, beam remnants, hadronization and multiple interactions. In section~\ref{sec:numerical_analysis}, we present the generated samples, parameter variations, and analysis observables used to probe the hard photon-induced system, charged-particle activity, rapidity gaps and remnant-sensitive final states. The main results are discussed in section~\ref{sec:results}, where we compare the three production modes and assess the impact of the shower setting variations. We conclude in section~\ref{sec:conc}, highlighting the implications for modelling exclusive and semi-exclusive photon-induced measurements and outlining the generator comparisons left for future work.

\section{Photon-Induced Production in Herwig 7}
\label{sec:pi-herwig}

Photon-induced production can be organised according to whether neither, one, or both incoming protons dissociate after emitting the photons that participate in the hard interaction. We denote the corresponding topologies by
\begin{equation}
pp \to p+X+p,\qquad
pp \to p+X+Y,\qquad
pp \to Y_1+X+Y_2 ,
\label{eq:photon-topologies}
\end{equation}
where $X$ is the central photon-induced system and $Y_i$ denotes a proton-dissociation system. The first topology corresponds to coherent, or elastic, photon emission from both incoming protons. The second and third correspond to single- and double-dissociative production, respectively.

At the level of the incoming photon densities, the cross section may be written schematically as a sum over elastic and inelastic photon sources,
\begin{equation}
d\sigma(pp\to X)
=
\sum_{a,b={\rm el,inel}}
\int dx_1 dx_2\,
f_{\gamma/p}^{a}(x_1,\mu_F^2)\,
f_{\gamma/p}^{b}(x_2,\mu_F^2)\,
d\hat{\sigma}_{\gamma\gamma\to X}(\hat{s},\mu_F^2),
\label{eq:photon-master}
\end{equation}
with $\hat{s}=x_1x_2s$. Here $s$ is the squared proton--proton centre-of-mass energy, $\hat{s}$ is the squared invariant mass of the partonic $\gamma\gamma$ system, and $x_1$ and $x_2$ are the longitudinal momentum fractions carried by the two photons. The indices $a,b={\rm el,inel}$ label whether the photon is emitted coherently from an intact proton or incoherently in association with proton dissociation. The functions $f_{\gamma/p}^{\rm el}$ and $f_{\gamma/p}^{\rm inel}$ denote the corresponding elastic photon flux and inelastic photon density, respectively, while $d\hat{\sigma}_{\gamma\gamma\to X}$ is the partonic cross section for producing the central system $X$ from the two incoming photons. The scale $\mu_F$ denotes the factorisation scale associated with the photon density; where this dependence is not essential, it will be suppressed below. The explicit dependence of the elastic flux on the photon virtuality and proton electromagnetic form factors is also not shown in eq.~\eqref{eq:photon-master}, since the expression is intended only to define the separation of event classes used in the generator setup.

The three event classes used in this study are obtained by retaining the corresponding terms in eq.~\eqref{eq:photon-master}. The coherent contribution contains elastic photon emission from both protons,
\begin{align}
d\sigma_{\rm coh}
&=
\int dx_1 dx_2\,
f_{\gamma/p}^{\rm el}(x_1,\mu_F^2)\,
f_{\gamma/p}^{\rm el}(x_2,\mu_F^2)\,
d\hat{\sigma}_{\gamma\gamma\to X}(\hat{s},\mu_F^2),
\label{eq:sigma-coh}
\end{align}
and therefore corresponds to the topology $pp\to p+X+p$. The single-dissociative contribution contains one elastic and one inelastic photon source,
\begin{align}
d\sigma_{\rm sD}
&=
\int dx_1 dx_2\,
\left[
f_{\gamma/p}^{\rm el}(x_1,\mu_F^2)\,
f_{\gamma/p}^{\rm inel}(x_2,\mu_F^2)
+
f_{\gamma/p}^{\rm inel}(x_1,\mu_F^2)\,
f_{\gamma/p}^{\rm el}(x_2,\mu_F^2)
\right]
d\hat{\sigma}_{\gamma\gamma\to X}(\hat{s},\mu_F^2),
\label{eq:sigma-sd}
\end{align}
The physical single-dissociative contribution contains both beam orientations, $pp\to p+X+Y$ and $pp\to Y+X+p$. In this study, we represent one orientation explicitly, with beam A inelastic and beam B elastic; the opposite orientation is obtained by interchanging the two identical proton beams. Finally, the double-dissociative contribution contains inelastic photon emission from both protons,
\begin{align}
d\sigma_{\rm dD}
&=
\int dx_1 dx_2\,
f_{\gamma/p}^{\rm inel}(x_1,\mu_F^2)\,
f_{\gamma/p}^{\rm inel}(x_2,\mu_F^2)\,
d\hat{\sigma}_{\gamma\gamma\to X}(\hat{s},\mu_F^2),
\label{eq:sigma-dd}
\end{align}
and corresponds to $pp\to Y_1+X+Y_2$. The coherent contribution is determined by the electromagnetic structure of the proton and is usually expressed in terms of elastic form factors or an equivalent-photon flux~\cite{Budnev:1975poe,Drees:1988pp}. By contrast, the inelastic contribution is associated with proton dissociation and can be described either through proton structure functions or through an inelastic photon PDF~\cite{Manohar:2016nzj,Manohar:2017eqh}.These expressions provide a compact definition of the elastic-inelastic decomposition implemented in the \textsf{Herwig~7} samples below.

The distinction between the three classes is therefore not only a matter of normalisation. In coherent production, the photon is emitted by the proton as a whole and the emitting proton remains intact. The final state contains the colour-singlet hard system $X$ and, up to additional soft interactions, no dissociative beam remnant. In single-dissociative production, one side of the event has this coherent structure, while the other contains an inelastic photon emission and a proton remnant. In double-dissociative production, both sides contain such remnant systems. The central matrix element may be the same in all three cases, but the particle-level event structure is not: the dissociative components can generate additional forward and central charged activity through remnant fragmentation, showering and hadronization. These effects are precisely those probed by rapidity-gap, track-veto and forward-activity observables.

In the \textsf{Herwig~7} input files the three event classes are defined by the photon source assigned to the two incoming proton beams. For the coherent sample, both incoming photons are taken from the elastic photon source,
\begin{verbatim}
set /Herwig/Particles/p+:PDF /Herwig/Partons/BudnevPDF
set /Herwig/Particles/pbar-:PDF /Herwig/Partons/BudnevPDF
set /Herwig/Partons/PPExtractor:FirstPDF /Herwig/Partons/BudnevPDF
set /Herwig/Partons/PPExtractor:SecondPDF /Herwig/Partons/BudnevPDF
set /Herwig/Shower/ShowerHandler:PDFA NULL
set /Herwig/Shower/ShowerHandler:PDFB NULL
\end{verbatim}
so that neither side is associated with an inelastic beam remnant. For the dissociative side we use the inelastic-only MSHT20qed photon PDF,
\begin{verbatim}
set /Herwig/Partons/HardLOPDF:PDFName MSHT20qed_nnlo_inelastic
set /Herwig/Partons/ShowerLOPDF:PDFName MSHT20qed_nnlo_inelastic
set /Herwig/Partons/RemnantPDF:PDFName MSHT20qed_nnlo_inelastic
set /Herwig/Partons/RemnantDecayer:AllowLeptons Yes
\end{verbatim}
In the single-dissociative setup used here, beam~A is inelastic and beam~B is elastic,
\begin{verbatim}
set /Herwig/Partons/PPExtractor:FirstPDF /Herwig/Partons/HardLOPDF
set /Herwig/Partons/PPExtractor:SecondPDF /Herwig/Partons/BudnevPDF
set /Herwig/Shower/ShowerHandler:PDFA /Herwig/Partons/ShowerLOPDF
set /Herwig/Shower/ShowerHandler:PDFB NULL
\end{verbatim}
whereas in the double-dissociative setup both beams use the inelastic photon PDF,
\begin{verbatim}
set /Herwig/Partons/PPExtractor:FirstPDF /Herwig/Partons/HardLOPDF
set /Herwig/Partons/PPExtractor:SecondPDF /Herwig/Partons/HardLOPDF
set /Herwig/Shower/ShowerHandler:PDFA /Herwig/Partons/ShowerLOPDF
set /Herwig/Shower/ShowerHandler:PDFB /Herwig/Partons/ShowerLOPDF
\end{verbatim}
The inelastic photon PDF is used instead of an inclusive QED photon PDF in order to keep the dissociative samples exclusive of the coherent contribution already described by the elastic photon flux, and to avoid mixing elastic and inelastic photon sources in the sample definition~\cite{Cridge:2021pxm}. The operational definition of the three samples is summarised in table~\ref{tab:sample-definitions}.

\begin{table}
\centering
\resizebox{0.95\textwidth}{!}{%
\begin{tabular}{c c c c c}
\toprule
Sample & Beam 1 photon source & Beam 2 photon source & Remnants & Internal matrix element \\
\midrule
coherent
& elastic
& elastic
& none
& \texttt{MEgg2ff} or \texttt{MEgg2WW}
\\
single-dissociative
& inelastic
& elastic
& one side
& \texttt{MEgg2ff} or \texttt{MEgg2WW}
\\
double-dissociative
& inelastic
& inelastic
& both sides
& \texttt{MEgg2ff} or \texttt{MEgg2WW}
\\
\bottomrule
\end{tabular}}
\caption{Operational definition of the coherent, single-dissociative and double-dissociative samples in the \textsf{Herwig~7} input files. Here ``elastic'' denotes \texttt{BudnevPDF}, while ``inelastic'' denotes \texttt{MSHT20qed\_nnlo\_inelastic}. The same internal matrix elements are used for all three photon-source configurations: \texttt{MEgg2ff} for $\gamma\gamma\to\ell^+\ell^-$ and \texttt{MEgg2WW} for $\gamma\gamma\to W^+W^-$.}
\label{tab:sample-definitions}
\end{table}

The hard subprocess is generated using the internal \textsf{Herwig~7} matrix elements for photon--photon scattering. For dilepton production, we use \texttt{MEgg2ff}, with separate electron and muon configurations,
\begin{verbatim}
cd /Herwig/MatrixElements

cp MEgg2ff MEgg2ee
cp gg2ffAmp gg2eeAmp
set MEgg2ee:Amplitude gg2eeAmp
set gg2eeAmp:Process Electron
insert SubProcess:MatrixElements 0 MEgg2ee

cp MEgg2ff MEgg2mm
cp gg2ffAmp gg2mmAmp
set MEgg2mm:Amplitude gg2mmAmp
set gg2mmAmp:Process Muon
insert SubProcess:MatrixElements 1 MEgg2mm
\end{verbatim}
This gives the tree-level QED processes $\gamma\gamma\to e^+e^-$ and $\gamma\gamma\to\mu^+\mu^-$. For diboson production, we use
\begin{verbatim}
cd /Herwig/MatrixElements
insert SubProcess:MatrixElements 0 MEgg2WW
set MEgg2WW:MassOption OffShell
\end{verbatim}
which describes $\gamma\gamma\to W^+W^-$. The leptonic final state $\ell^+\ell^-\nu\bar{\nu}$ is obtained by restricting the subsequent \textsf{Herwig} decays of the two $W$ bosons to electron and muon modes. In both channels, the same internal hard matrix elements are used for the coherent, single-dissociative and double-dissociative samples; the distinction between these samples enters through the photon sources and the corresponding beam-remnant structure rather than through the central $\gamma\gamma$ amplitude.

The internal \textsf{Herwig~7} matrix elements are used in order to keep the photon-source and remnant modelling under direct control. For the two processes considered here, the required hard amplitudes are simple tree-level $\gamma\gamma$ reactions, so the internal implementations provide a clean reference without introducing additional dependencies from external matrix-element generation, event-file interfaces, matching settings or process-dependent phase-space mappings. This is particularly important for the present study, since the main question is not the structure of the hard $\gamma\gamma$ amplitude itself, but how the coherent and dissociative photon sources are connected to showering, beam remnants, hadronization and particle-level exclusivity observables. 

The perturbative shower content is controlled through the \texttt{Interactions} option of the \textsf{Herwig~7} shower handler. In the central samples we use
\begin{verbatim}
set /Herwig/Shower/ShowerHandler:Interactions ALL
\end{verbatim}
which corresponds to the combined QCD+QED+EW angular-ordered shower. This option is the relevant default for the present study, since photon-induced events are not purely electromagnetic once proton dissociation is included. The hard process is electroweak, but the dissociative beam remnants are coloured and can generate QCD radiation and hadronic activity, while QED and EW radiation can affect the leptonic and diboson final states. The \texttt{ALL} setting therefore provides a unified treatment of QCD, QED and EW emissions in the same shower evolution, building on the implementation of the angular-ordered EW shower in \textsf{Herwig~7} and its subsequent developments~\cite{Masouminia:2021kne,Darvishi:2021het,Lee:2023hef,Kim:2026mfk,Masouminia:2026udz}. For comparison, we also consider restricted shower settings, in which only QED radiation is retained. The QED-only comparison is useful diagnostically, but it removes both QCD and EW shower emissions, so differences from the full shower cannot be attributed solely to QCD activity associated with dissociative remnants. They should not, however, be interpreted as conventional perturbative uncertainty variations: switching from \texttt{ALL} to a restricted shower changes the physical content of the simulation. This distinction is especially important for the single- and double-dissociative samples, where the modelling of charged activity, rapidity gaps and forward-remnant observables depends directly on the showering and fragmentation of the remnant system.

The remnant treatment is controlled by the PDF assignment on the dissociative side and by the corresponding remnant machinery, including the remnant PDF and remnant-decay settings discussed above. Hadronization is performed with the standard \textsf{Herwig~7} cluster model, so that any coloured remnant system produced by an inelastic photon emission is evolved into hadrons before the particle-level analysis. Multiple partonic interactions are disabled throughout the present study,
\begin{verbatim}
set /Herwig/Shower/ShowerHandler:MPIHandler NULL
\end{verbatim}
so that additional soft proton-proton activity is not included. Consequently, MPI-related tune parameters are not physically active in the samples considered here. The physically relevant variations are instead those that affect the shower and remnant evolution, such as the \texttt{Interactions} setting, the shower cut-off scale, and the QCD coupling variations for initial- and final-state radiation. In coherent production, these QCD variations are expected to have little or no effect, because there is no dissociative coloured remnant. In the dissociative samples, they can affect the event only through the coloured remnant and its subsequent shower and hadronization, rather than through the central $\gamma\gamma$ matrix element itself.

Before turning to the distributions, it is useful to state the main technical limitations of the setup. First, the coherent and dissociative components are defined operationally through the photon sources assigned in the \textsf{Herwig} input files: the elastic flux for intact-proton emission and the inelastic photon PDF for proton dissociation. This gives a clean separation between elastic and inelastic photon sources, but it does not constitute a complete theory prediction for exclusive photon-induced production, since soft-survival effects and additional soft proton-proton interactions are not modelled. In particular, MPI is disabled throughout the study, so the results should be interpreted as isolating the shower, remnant and hadronization structure associated with the photon-induced event itself, rather than as fully inclusive predictions for experimental exclusivity efficiencies.

Second, the dissociative samples rely on the event-generator treatment of the beam remnant following the extraction of an inelastic photon from the proton. The remnant fragmentation, possible leading-baryon production and the migration of remnant particles into the detector acceptance are therefore model-dependent. This is precisely one of the effects we wish to study, but it also means that remnant-sensitive observables should not be interpreted as universal predictions independent of the \textsf{Herwig} remnant and hadronization model.

Third, the hard processes considered here are generated with the internal \textsf{Herwig~7} matrix elements for $\gamma\gamma\to\ell^+\ell^-$ and $\gamma\gamma\to W^+W^-$. For the diboson channel, this means that the internal setup describes the resonant $\gamma\gamma\to W^+W^-$ topology followed by $W$ decays, rather than the full four-fermion matrix element. 
This provides a controlled baseline for comparing coherent, single-dissociative and double-dissociative configurations, but it does not replace a full comparison to alternative matrix-element and shower combinations. 
A systematic comparison with the \textsf{Matchbox}/\textsf{MadGraph} setup and \textsf{MadGraph} interfaced to \textsf{Pythia}, and with other photon-induced generators, is left for future work.

Finally, no detector simulation, pile-up modelling or experimental unfolding is included. The observables studied below should therefore be read as particle-level diagnostics of the generator modelling, especially of the separation between hard-system kinematics and remnant-induced event activity. This is sufficient for establishing a Herwig-based baseline, but not for deriving final experimental correction factors or survival probabilities.

\section{Numerical Analysis}
\label{sec:numerical_analysis}

The numerical analysis considers two photon-induced final states, $\gamma\gamma\to\ell^+\ell^-$ and $\gamma\gamma\to W^+W^-\to\ell^+\ell^-\nu\bar{\nu}$, for coherent, single-dissociative and double-dissociative production. The central samples use the internal \textsf{Herwig~7} matrix elements and the default \texttt{QCD+QED+EW} shower setup described in section~\ref{sec:pi-herwig}. We additionally consider a QED-only shower, variations of the shower transverse-momentum cut-off, and variations of the QCD coupling used for initial- and final-state radiation. The QED-only setup is used as a diagnostic variation rather than as a perturbative uncertainty. The particle-level analysis is performed with \textsf{Rivet}~\cite{Bierlich:2024Rivet}, with histogramming and statistical data handling provided by \textsf{YODA}~\cite{Buckley:2025YODA}. Electrons and muons are dressed with photons within $\Delta R=0.1$ and required to satisfy $p_T>10~\mathrm{GeV}$ and $|\eta|<2.5$. Additional charged particles are selected with $p_T>500~\mathrm{MeV}$ and $|\eta|<2.5$, while energetic final-state protons are identified by $|p_z|>5~\mathrm{TeV}$. 

The hard photon-induced system is characterised by the invariant mass and rapidity of the $\gamma\gamma$ system,
\begin{equation}
M_{\gamma\gamma}=\sqrt{x_1x_2s}, \qquad
y_{\gamma\gamma}=\frac{1}{2}\log\frac{x_1}{x_2},
\end{equation}
together with the photon momentum fractions, shown as $\log_{10}x_\gamma$ for photons travelling in the positive- and negative-$z$ directions. For the dilepton channel we additionally consider $m_{\ell\ell}$, $p_{T,\ell\ell}$, $\Delta\phi_{\ell\ell}$ and the individual lepton kinematics. These distributions provide direct checks of the central photon-induced process and are contrasted with observables sensitive to proton dissociation.

Exclusivity and dissociative activity are probed using charged particles other than the signal leptons. We consider the non-lepton charged-particle multiplicity and pseudorapidity distribution, the pseudorapidity gap to the central system, the scalar charged-particle $p_T$ sum and the maximum charged-particle $p_T$. These observables characterise the amount, location and hardness of additional activity and are directly related to experimental track-veto and rapidity-gap selections. The forward event structure is probed using energetic final-state protons. Their occurrence in the positive- and negative-$z$ directions, together with their longitudinal and transverse momenta, provides information on the intact and dissociative proton configurations. In dissociative samples, such protons may originate from remnant fragmentation and should therefore not necessarily be interpreted as elastically scattered protons.

\section{Results and Discussion}
\label{sec:results}

\begin{figure}
    \centering
    \includegraphics[width=0.49\textwidth]{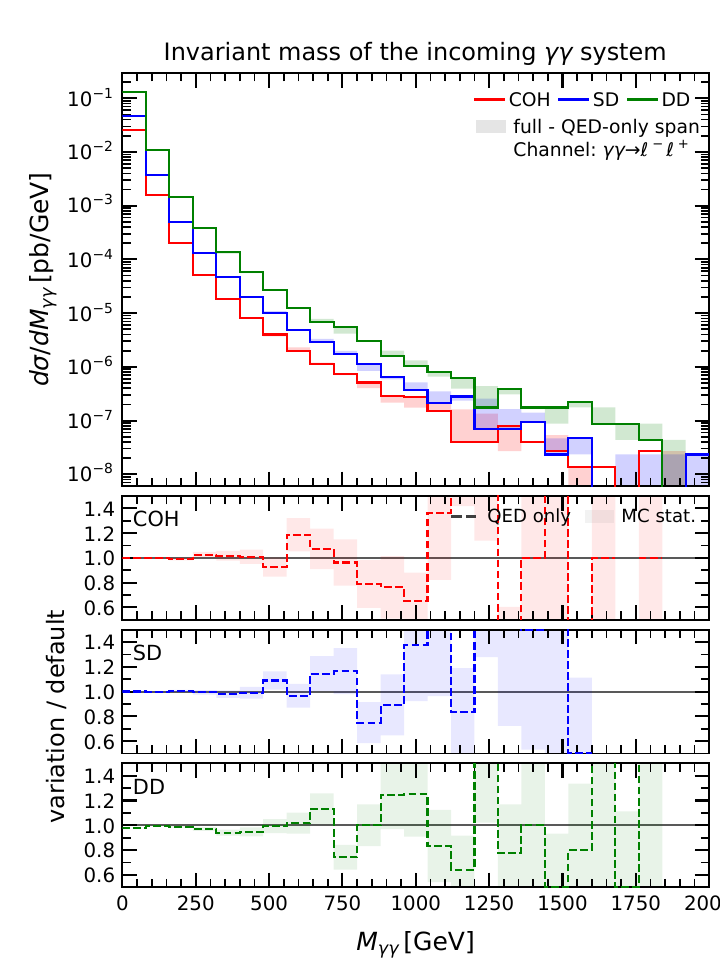}
    \includegraphics[width=0.49\textwidth]{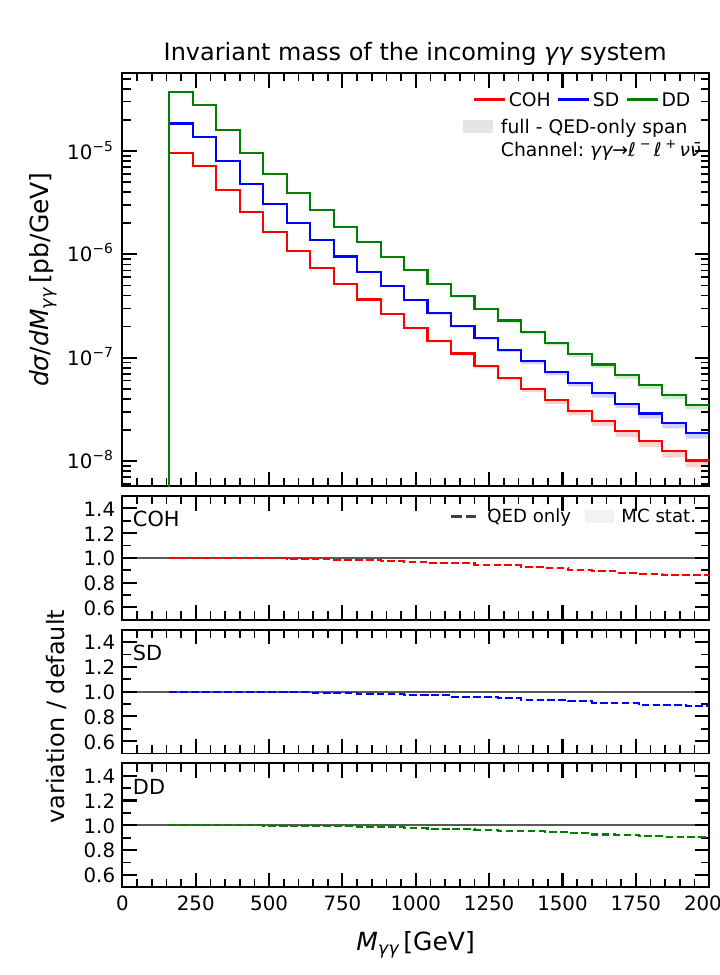}
    \caption{Differential cross section as a function of the invariant mass $M_{\gamma\gamma}$ of the incoming photon pair in $\gamma\gamma\to\ell^+\ell^-$ (left) and $\gamma\gamma\to W^+W^-\to\ell^+\ell^-\nu\bar{\nu}$ (right). The solid red, blue and green curves show the default full-shower predictions for coherent (COH), single-dissociative (SD) and double-dissociative (DD) photon emission, respectively; the SD sample represents one beam orientation. The coloured shading spans the full and QED-only shower predictions. The three lower panels show the QED-only prediction divided by the full prediction for COH, SD and DD, respectively, with shaded Monte Carlo statistical uncertainties.}
    \label{fig:digamma-mass-xsec}
\end{figure}

\begin{figure}
    \centering
    \includegraphics[width=0.49\textwidth]{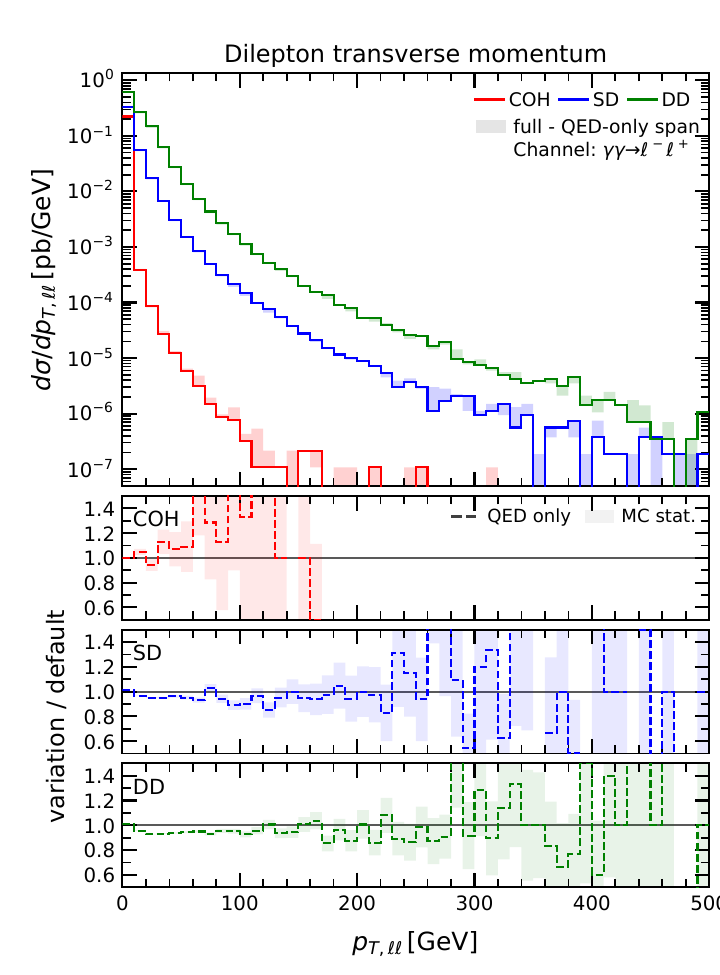}
    \includegraphics[width=0.49\textwidth]{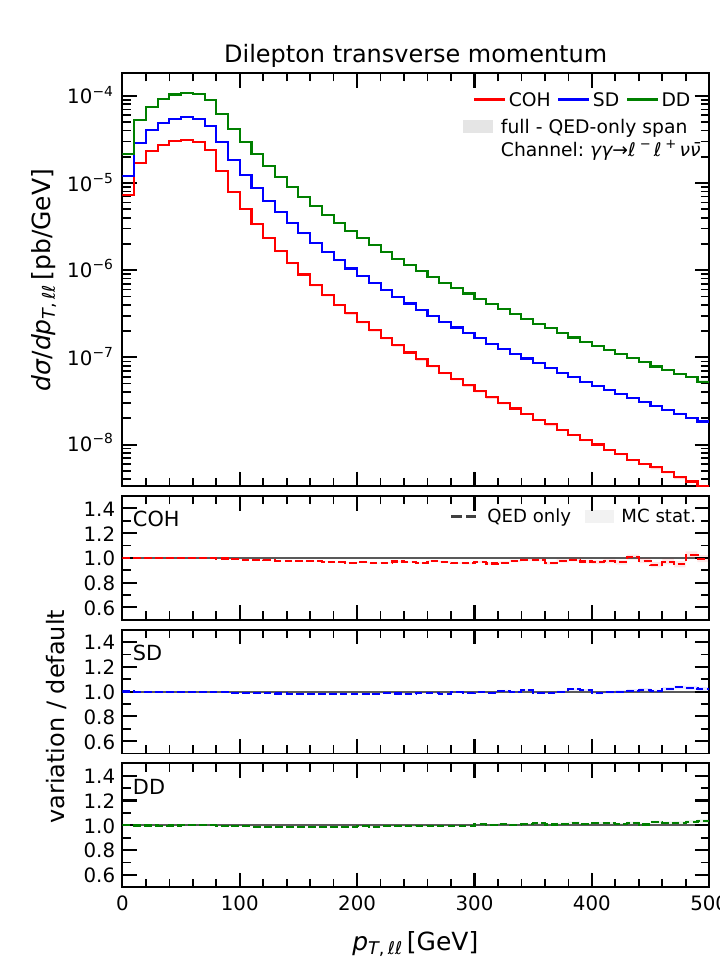}
    \caption{Differential cross section as a function of the dilepton transverse momentum $p_{T,\ell\ell}$ in $\gamma\gamma\to\ell^+\ell^-$ (left) and $\gamma\gamma\to W^+W^-\to\ell^+\ell^-\nu\bar{\nu}$ (right). The notation is identical to that of Fig.~\ref{fig:digamma-mass-xsec}.}
    \label{fig:ptll-xsec}
\end{figure}

Within each channel, Figs.~\ref{fig:digamma-mass-xsec} and \ref{fig:ptll-xsec} show that using the same internal matrix element for the three photon-source configurations does not make their hard-system distributions identical. The $M_{\gamma\gamma}$ spectra fall steeply in all three configurations but differ in absolute rate. The $p_{T,\ell\ell}$ spectra show a stronger shape dependence in direct dilepton production: coherent events are concentrated near $p_{T,\ell\ell}=0$, whereas the dissociative samples extend to substantially larger transverse momenta. The diboson dilepton momentum is also shaped by the subsequent $W$ decays. Changing from the full to the QED-only shower generally has a smaller effect in the statistically well-populated regions of these hard-system distributions; fluctuations in the dilepton tails should not be overinterpreted.

\begin{figure}
    \centering
    \includegraphics[width=0.49\textwidth]{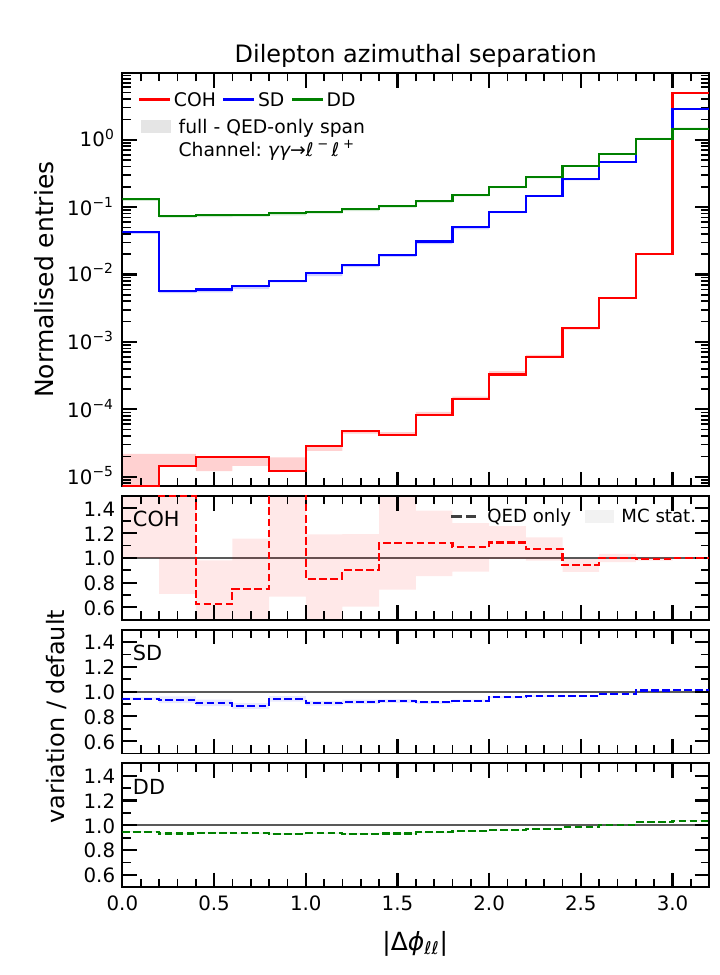}
    \includegraphics[width=0.49\textwidth]{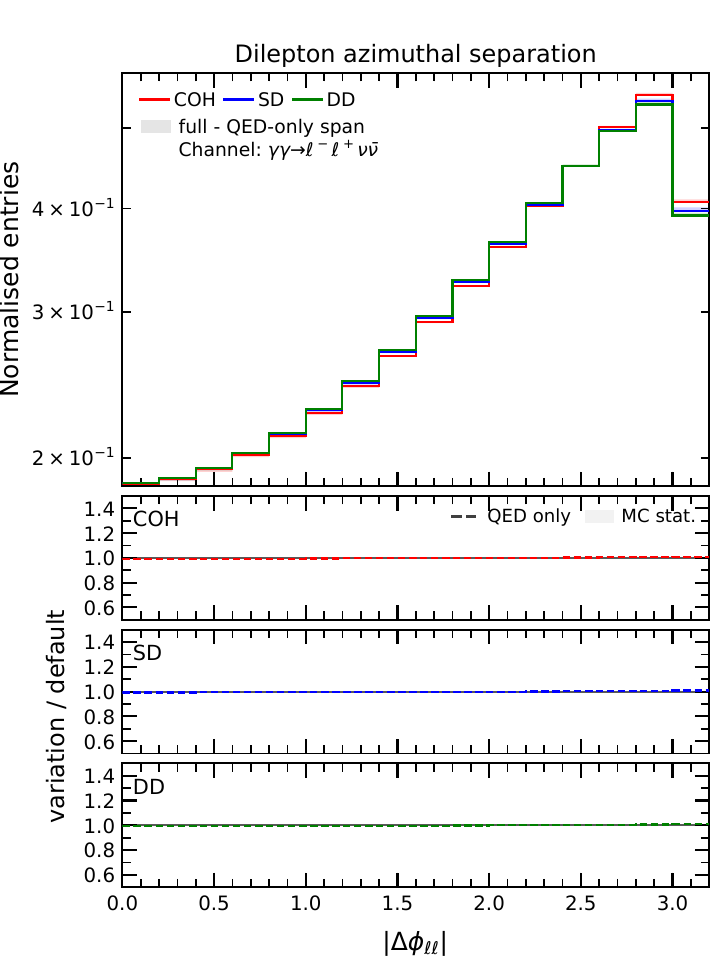}
    \caption{Distribution of the azimuthal separation $\Delta\phi_{\ell\ell}$ between the charged leptons in $\gamma\gamma\to\ell^+\ell^-$ (left) and $\gamma\gamma\to W^+W^-\to\ell^+\ell^-\nu\bar{\nu}$ (right). The notation is identical to that of Fig.~\ref{fig:digamma-mass-xsec}.}
    \label{fig:dphill}
\end{figure}

The visible lepton observables also distinguish the two processes. Direct $\gamma\gamma\to\ell^+\ell^-$ production has a pronounced back-to-back azimuthal correlation, whereas the charged leptons in $\gamma\gamma\to W^+W^-\to\ell^+\ell^-\nu\bar{\nu}$ arise from separate $W$ decays and the neutrinos carry away momentum, broadening the $p_{T,\ell\ell}$ distribution and weakening that correlation (Figs.~\ref{fig:ptll-xsec} and \ref{fig:dphill}). The direct dilepton channel therefore provides an internal check of how the configured photon sources affect a simple hard process, although these patterns do not independently validate the absolute cross sections.

\begin{figure}
    \centering
    \includegraphics[width=0.49\textwidth]{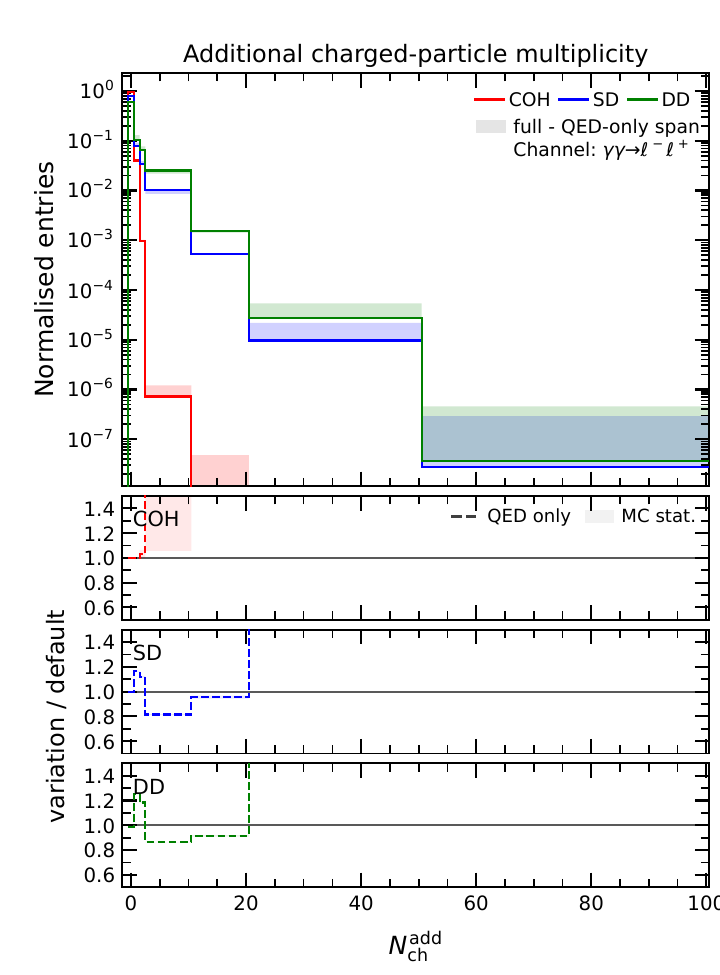}
    \includegraphics[width=0.49\textwidth]{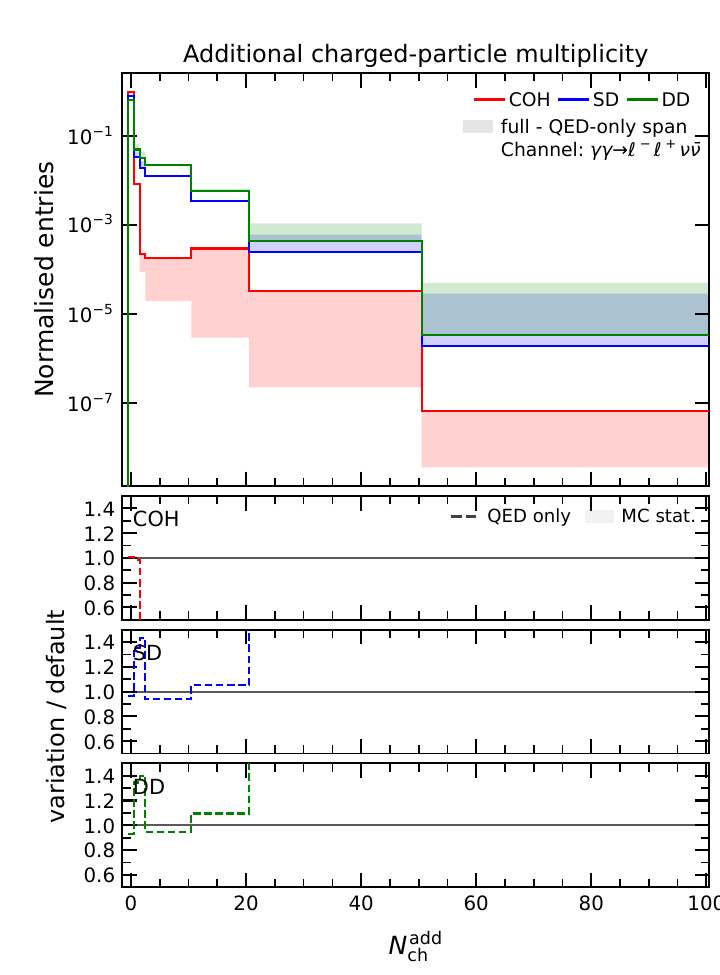}
    \caption{Event-level multiplicity of additional charged particles in $\gamma\gamma\to\ell^+\ell^-$ (left) and $\gamma\gamma\to W^+W^-\to\ell^+\ell^-\nu\bar{\nu}$ (right). The zero-multiplicity bin contains events with no selected additional charged particles. The notation is identical to that of Fig.~\ref{fig:digamma-mass-xsec}.}
    \label{fig:nlch-mult}
\end{figure}

\begin{figure}
    \centering
    \includegraphics[width=0.49\textwidth]{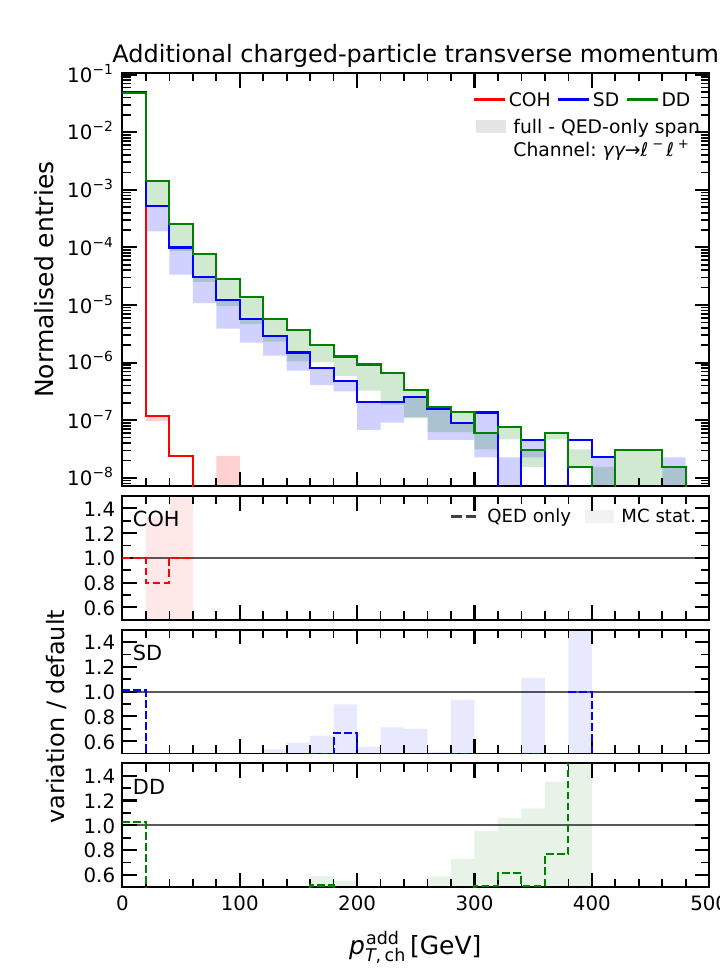}
    \includegraphics[width=0.49\textwidth]{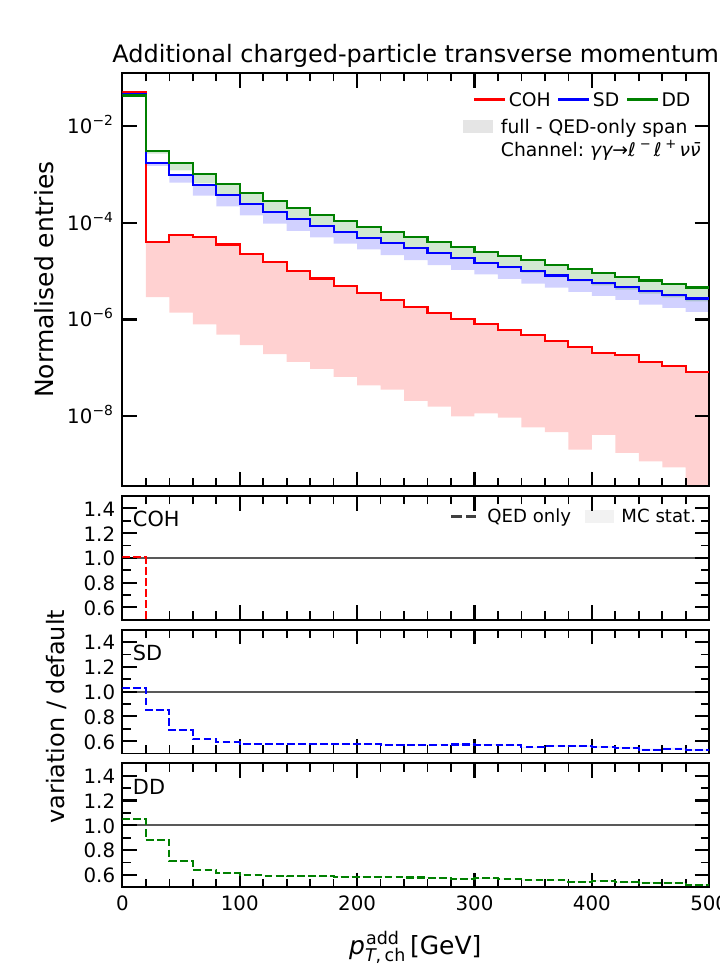}
    \caption{Event-level scalar transverse-momentum sum $\sum_{\mathrm{additional\ ch.}}p_T$ in $\gamma\gamma\to\ell^+\ell^-$ (left) and $\gamma\gamma\to W^+W^-\to\ell^+\ell^-\nu\bar{\nu}$ (right). Events with no selected additional charged particles have $\sum p_T=0$. The notation is identical to that of Fig.~\ref{fig:digamma-mass-xsec}.}
    \label{fig:nlch-ptsum}
\end{figure}

\begin{figure}
    \centering
    \includegraphics[width=0.49\textwidth]{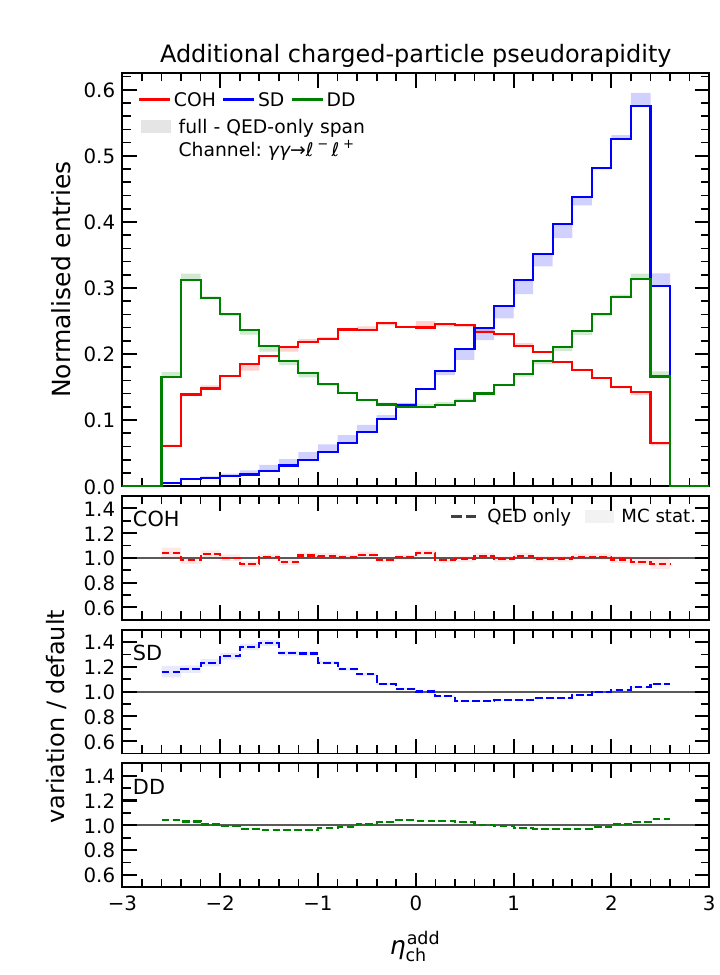}
    \includegraphics[width=0.49\textwidth]{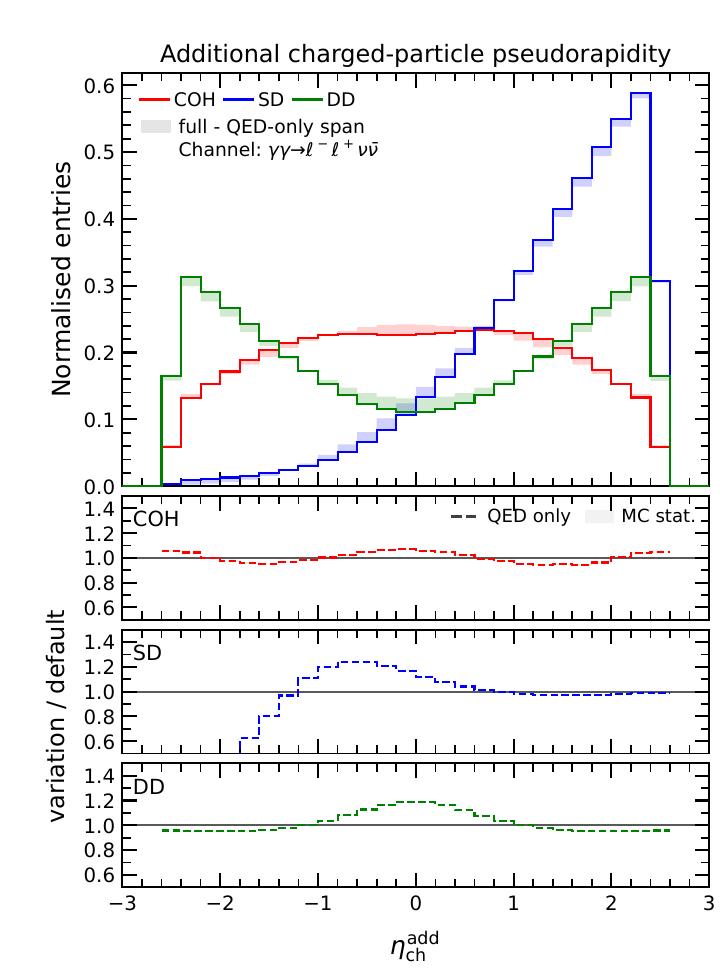}
    \caption{Pseudorapidity distribution of additional charged particles in $\gamma\gamma\to\ell^+\ell^-$ (left) and $\gamma\gamma\to W^+W^-\to\ell^+\ell^-\nu\bar{\nu}$ (right). Each selected additional charged particle contributes an entry. The notation is identical to that of Fig.~\ref{fig:digamma-mass-xsec}.}
    \label{fig:nlch-eta}
\end{figure}

The largest differences between the photon-source configurations in both channels occur in observables sensitive to proton dissociation. Coherent events are concentrated in the zero-additional-track and low-activity regions, whereas the SD and DD samples populate higher charged-particle multiplicities and larger scalar $p_T$ sums (Figs.~\ref{fig:nlch-mult} and \ref{fig:nlch-ptsum}). Their charged-particle pseudorapidity distributions also differ: the generated SD sample is asymmetric because it contains only one beam orientation, while DD produces activity on both sides (Fig.~\ref{fig:nlch-eta}). The diboson channel thus shows the same qualitative pattern of accompanying activity as the direct dilepton channel, despite their different visible-lepton kinematics. The finite-gap cross sections provide another view of this activity, but their plotted range excludes events with no selected additional charged particle; those events appear in the zero-multiplicity bin of Fig.~\ref{fig:nlch-mult}.

\begin{figure}
    \centering
    \includegraphics[width=0.49\textwidth]{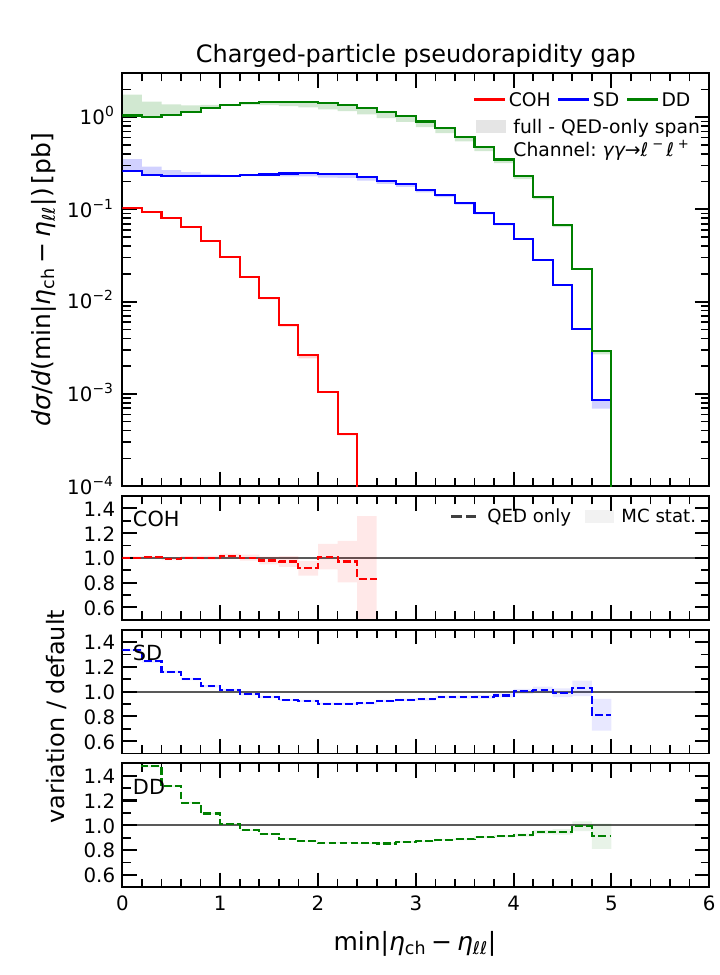}
    \includegraphics[width=0.49\textwidth]{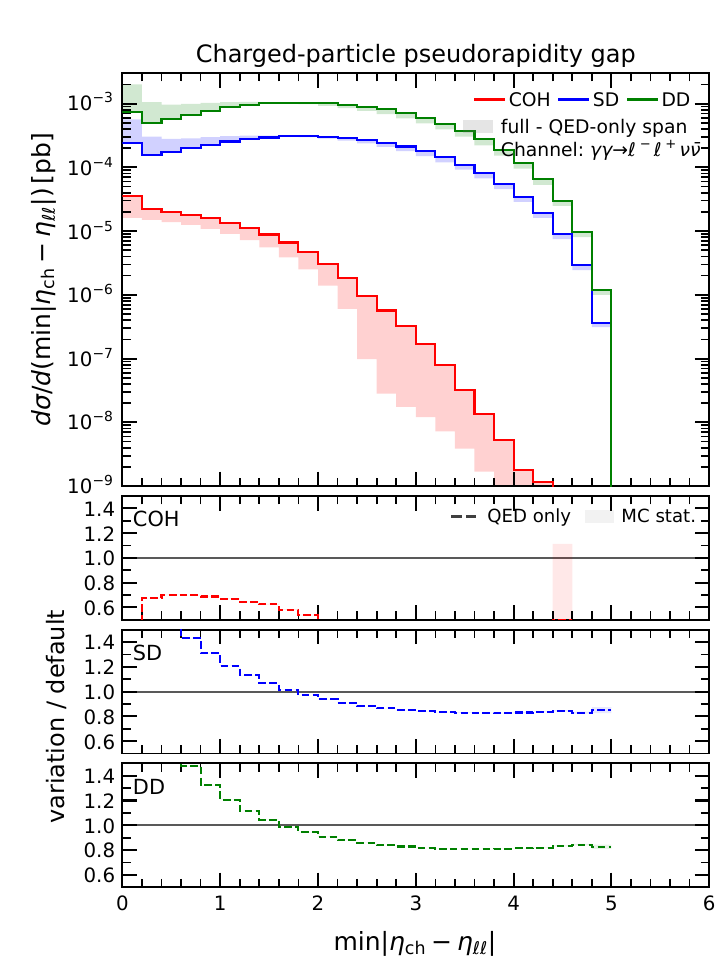}
    \caption{Differential cross section as a function of the minimum pseudorapidity separation $\Delta\eta_{\min}=\min_i|\eta_i-(\eta_{\ell^+}+\eta_{\ell^-})/2|$, where $i$ runs over selected additional charged particles, in $\gamma\gamma\to\ell^+\ell^-$ (left) and $\gamma\gamma\to W^+W^-\to\ell^+\ell^-\nu\bar{\nu}$ (right). Events with no selected additional charged particle lie outside the plotted range. The notation is identical to that of Fig.~\ref{fig:digamma-mass-xsec}.}
    \label{fig:nlch-etagap-xsec}
\end{figure}

\begin{figure}
    \centering
    \includegraphics[width=0.49\textwidth]{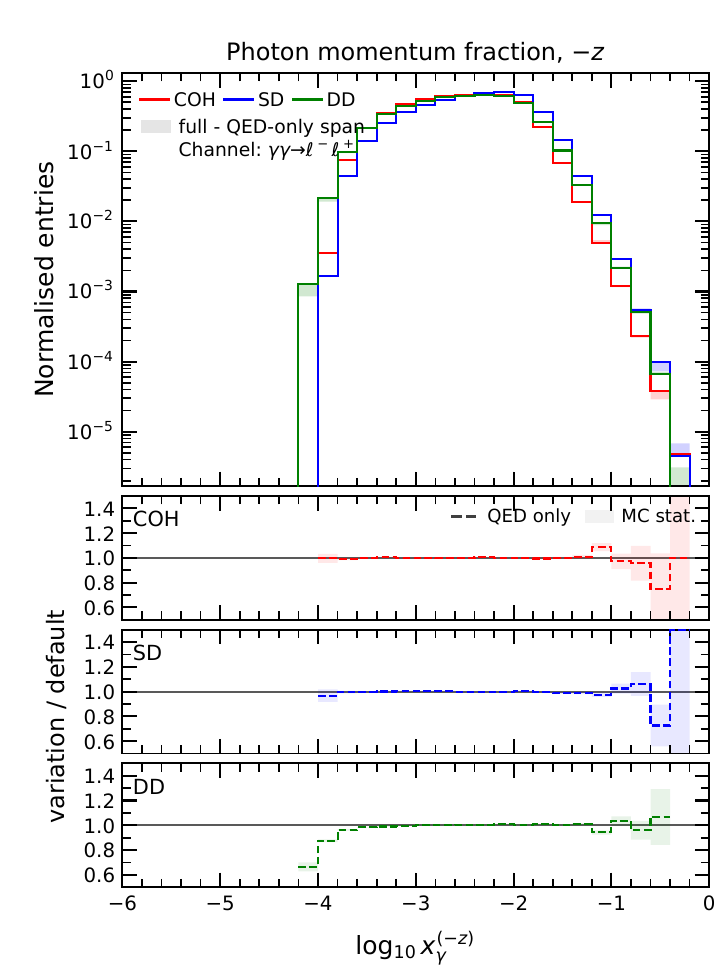}
    \includegraphics[width=0.49\textwidth]{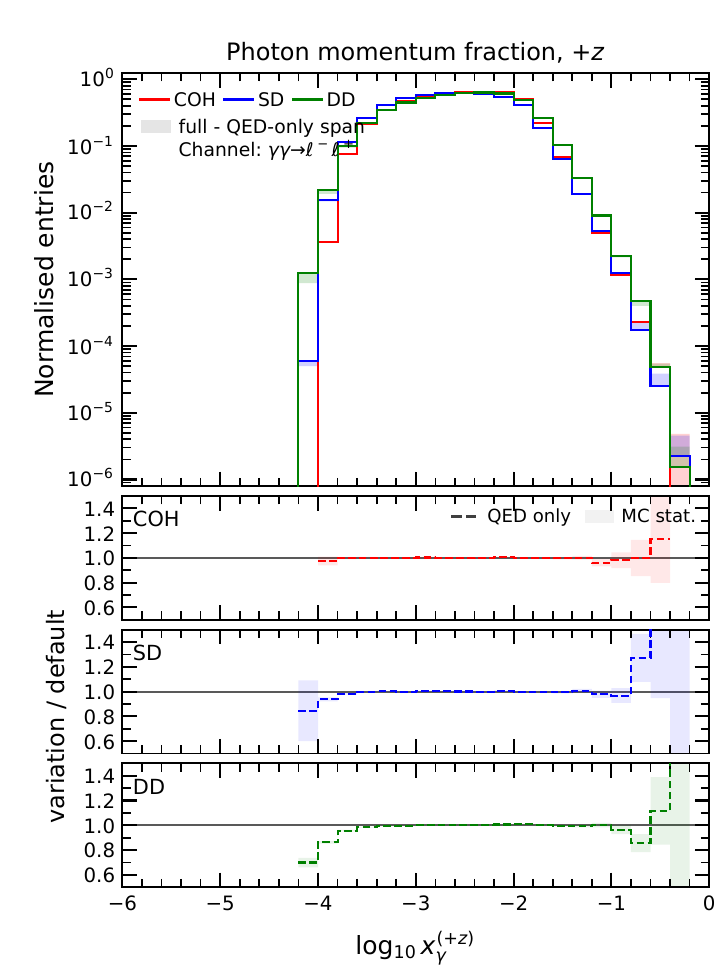}
    \caption{Distribution of $\log_{10}x_\gamma$, where $x_\gamma=(E_\gamma+|p_{z,\gamma}|)/\sqrt{s}$, for incoming photons travelling in the negative-$z$ (left) and positive-$z$ (right) directions in $\gamma\gamma\to\ell^+\ell^-$ production. The notation is identical to that of Fig.~\ref{fig:digamma-mass-xsec}.}
    \label{fig:ll-xgamma}
\end{figure}

\begin{figure}
    \centering
    \includegraphics[width=0.49\textwidth]{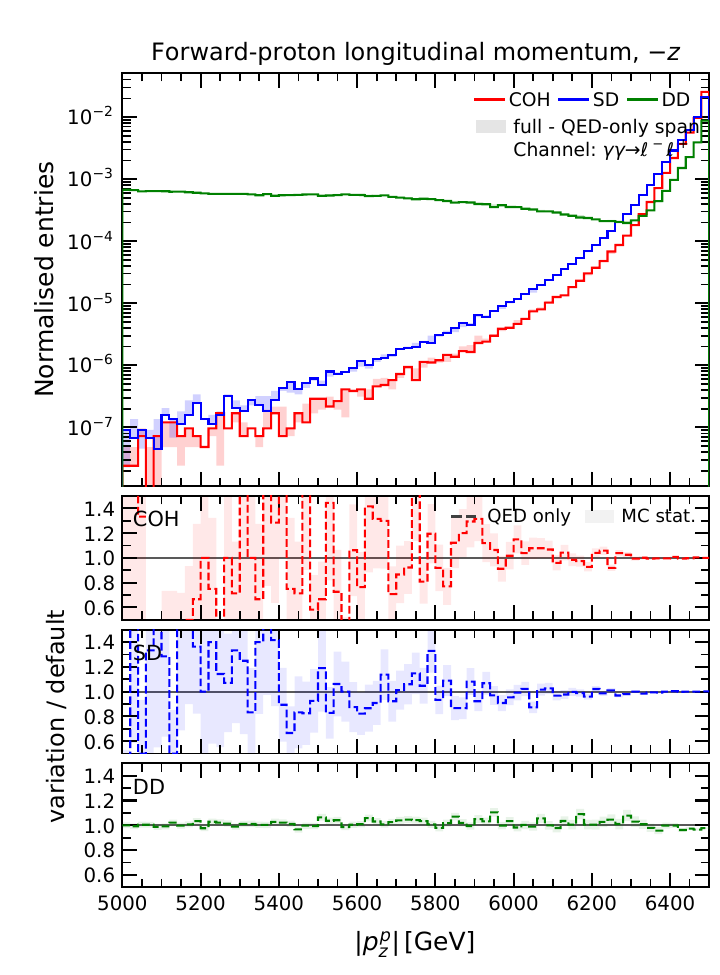}
    \includegraphics[width=0.49\textwidth]{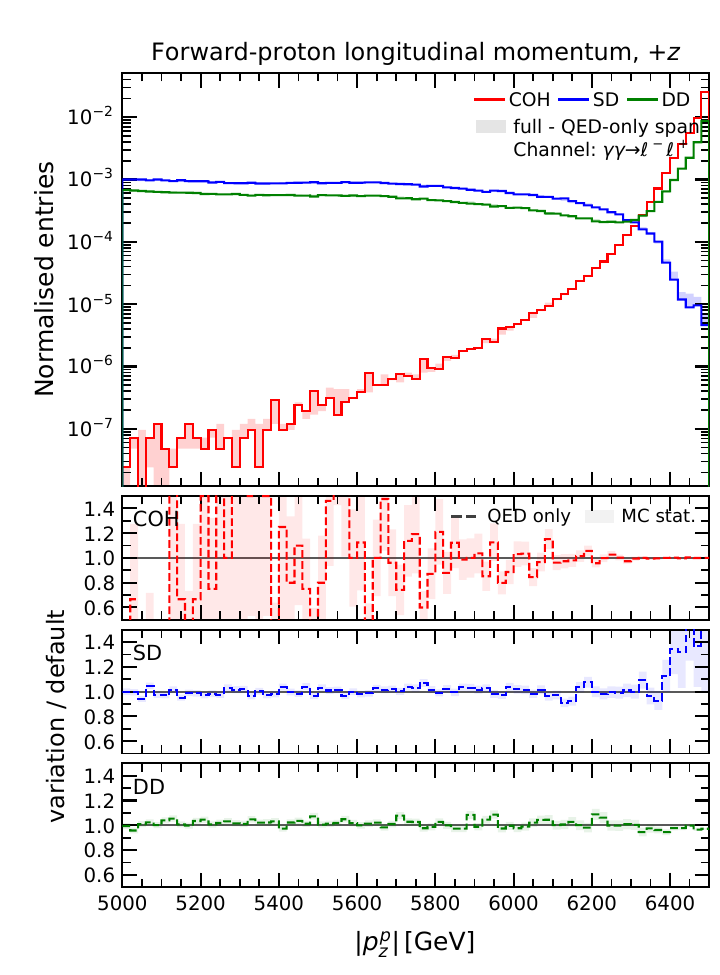}
    \caption{Distribution of $|p_z|$ for final-state protons with $|p_z|>5~\mathrm{TeV}$ travelling in the negative-$z$ (left) and positive-$z$ (right) directions in $\gamma\gamma\to\ell^+\ell^-$ production. The two distributions are normalised separately and therefore compare momentum shapes rather than proton yields. Protons in dissociative events may arise from remnant fragmentation. The notation is identical to that of Fig.~\ref{fig:digamma-mass-xsec}.}
    \label{fig:proton-pz}
\end{figure}

The side-resolved photon momentum fractions and forward-proton momentum distributions provide a further check of the beam assignments (Figs.~\ref{fig:ll-xgamma} and \ref{fig:proton-pz}). COH and DD have approximately similar shapes in the two beam directions, while the generated SD sample shows beam-dependent behaviour, particularly in the forward-proton shapes. This reflects the chosen SD orientation rather than a physical forward--backward asymmetry of the complete SD contribution. Forward protons on a dissociative side may also originate from remnant fragmentation rather than from an intact beam proton.

\begin{figure}
    \centering
    \includegraphics[width=0.8\textwidth]{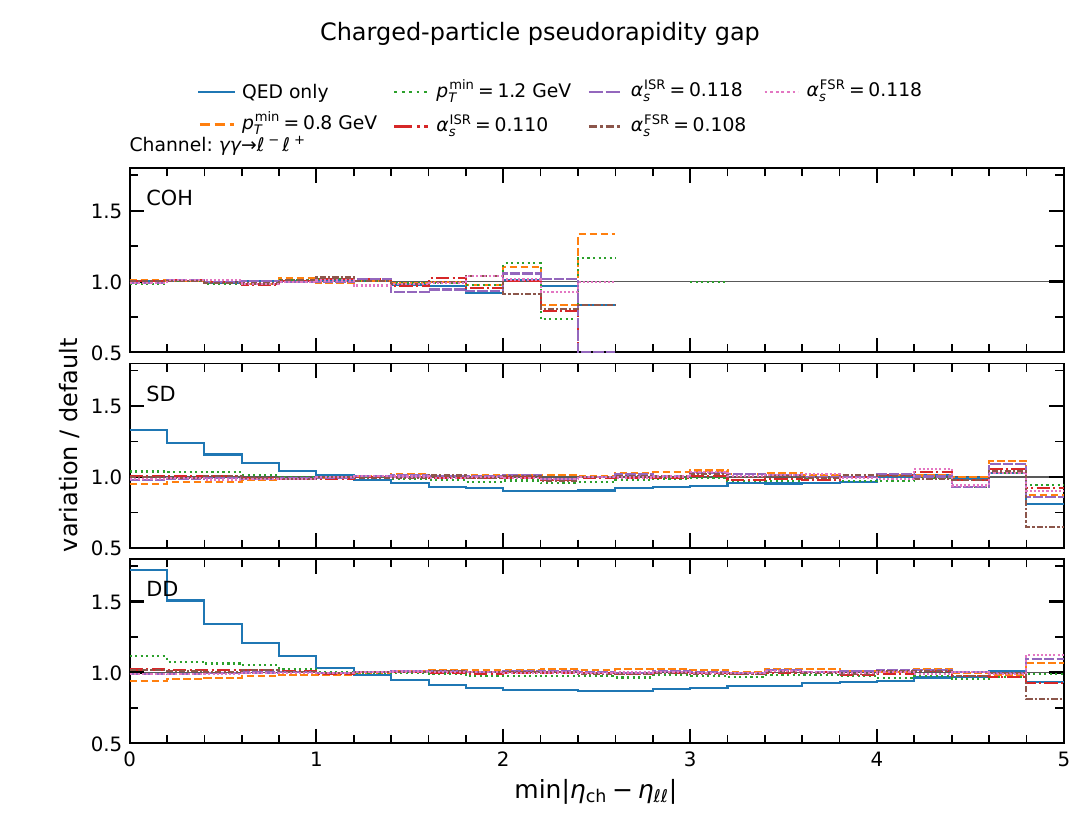}\\
    \includegraphics[width=0.8\textwidth]{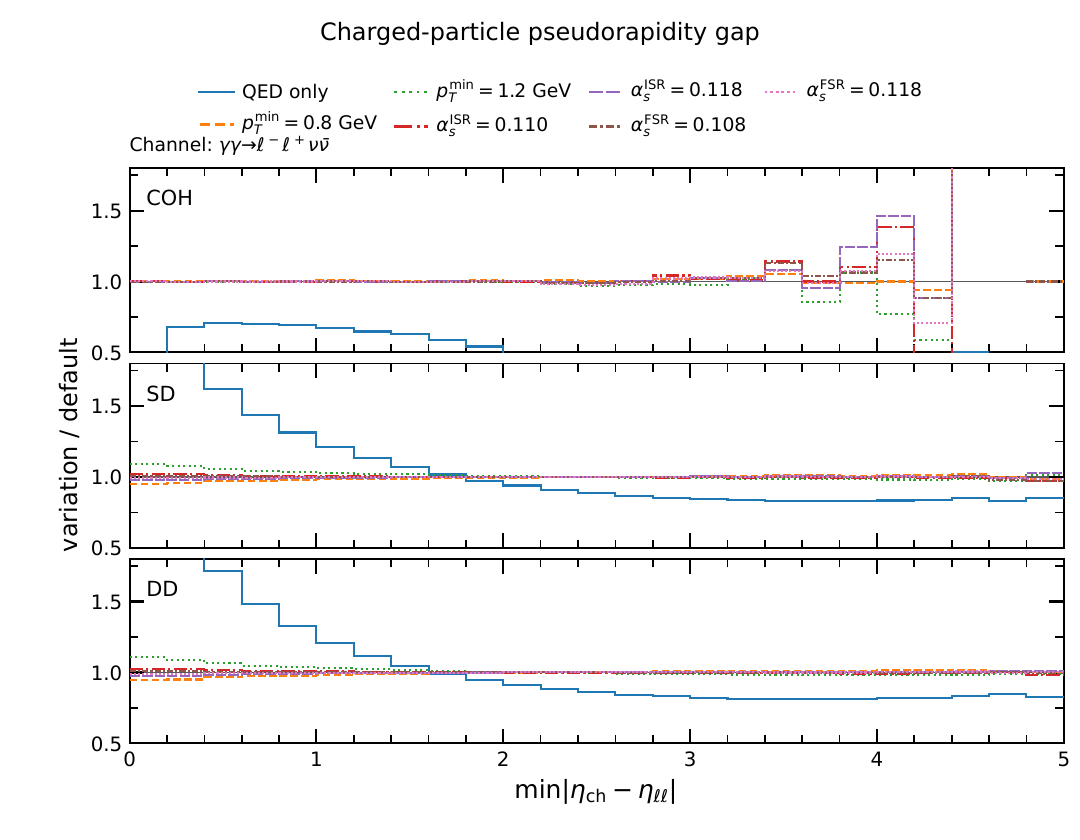}
    \caption{Ratios of the normalised charged-particle pseudorapidity-gap distributions to the default full QCD+QED+EW shower prediction in $\gamma\gamma\to\ell^+\ell^-$ (top) and $\gamma\gamma\to W^+W^-\to\ell^+\ell^-\nu\bar{\nu}$ (bottom). The three panels in each plot show coherent (COH), single-dissociative (SD) and double-dissociative (DD) production from top to bottom. The curves compare the QED-only shower, two shower $p_T$ cut-offs and variations of the ISR and FSR QCD couplings; the horizontal line at one marks the default prediction. Events with no selected additional charged particle lie outside the plotted gap range.}
    \label{fig:all-variations-etagap}
\end{figure}

Switching from the full QCD+QED+EW shower to QED-only produces the largest visible effect among the settings studied. It alters the additional charged-particle multiplicity, pseudorapidity and scalar $p_T$ sum, with clear effects in the dissociative samples (Figs.~\ref{fig:nlch-mult}, \ref{fig:nlch-eta} and \ref{fig:nlch-ptsum}). In the finite-gap ratios of Fig.~\ref{fig:all-variations-etagap}, the QED-only prediction enhances the smallest gaps and tends to suppress intermediate gaps for SD and DD. The coherent QED-only gap ratio is near unity in direct dilepton production but falls below unity over much of the displayed diboson range. This comparison changes the shower content and should be read as a diagnostic, not as an uncertainty band.

The ISR and FSR $\alpha_s$ variations remain close to unity over the well-populated gap range, including for coherent production, as expected for a configuration without a coloured beam remnant or coloured hard final state. Varying the shower cut-off from its nominal value of $p_T^{\min}=0.9565~\mathrm{GeV}$ to $0.8$ or $1.2~\mathrm{GeV}$ also gives mostly modest changes in that range. These comparisons vary shower settings rather than the remnant model itself. Because Fig.~\ref{fig:all-variations-etagap} has no MC statistical bands, the small shifts and larger excursions in sparsely populated bins do not establish a statistically significant $\alpha_s$ or cut-off dependence.

These results provide a particle-level \textsf{Herwig~7} study of the specified photon sources and shower settings rather than a complete prediction for experimental exclusive-event rates. MPI, soft-survival effects, pile-up and detector response are absent. The dissociative samples use one inelastic photon PDF and one remnant treatment, with no direct remnant-model variation. The generated SD sample contains only one beam orientation, so its absolute rate and directional distributions must be interpreted accordingly. Statistical fluctuations further limit conclusions in sparsely populated tails.

\section{Conclusions}
\label{sec:conc}

We have established a particle-level \textsf{Herwig~7} baseline for coherent, single-dissociative and double-dissociative photon-induced production in the dilepton and leptonic diboson channels. Using the same internal matrix element for the three photon sources within each channel makes their differences in incoming-photon kinematics and accompanying event activity directly visible. The sources differ in absolute rate and, particularly for direct dilepton production, in the dilepton transverse-momentum spectrum. Across both channels, the clearest distinction is the additional charged activity associated with proton dissociation. The single-dissociative results presented here represent one beam orientation.

Experimental studies should consider central-system kinematics alongside the fraction of events with no additional tracks, the multiplicity, pseudorapidity and scalar transverse-momentum sum of additional charged particles, the finite-gap distribution, and forward-proton observables. These quantities probe different aspects of the event selection: in particular, the plotted finite-gap distributions exclude events with no selected additional charged particle, and a forward proton in a dissociative event need not be an intact beam proton.

Among the shower settings examined, switching from the full QCD+QED+EW shower to QED-only gives the largest visible changes in charged activity and gap observables. The studied ISR and FSR $\alpha_s$ and shower cut-off variations generally produce smaller changes in well-populated regions. This makes the modelling of accompanying activity particularly important for exclusivity selections, but the present study cannot establish a quantitative ranking of theoretical uncertainties: the photon densities and remnant model were not varied, the QED-only comparison also removes EW shower emissions, and MPI and soft-survival effects were not included.

The internal matrix elements therefore remain the central prediction for this study. 
A \textsf{Matchbox}/\textsf{MadGraph} cross-check should first align the hard processes, photon sources, scales and selections and compare them before showering. Comparisons including \textsf{MadGraph} interfaced to \textsf{Pythia}, other generators and dedicated remnant and survival studies could then test the robustness of the particle-level predictions, with sufficient statistics and MC uncertainties shown on the ratios.

\section*{Acknowledgements}
\noindent We thank our fellow \textsf{Herwig~7} authors for useful discussions. We thank M.L.~Mangano, S.~Pl\"atzer, C.~G\"utschow, F.~Krauss and J.~Butterworth for their support of MCnet and its short-term studentship scheme. LG acknowledges support from the MCnet short-term studentship scheme and the LHC Physics Centre at CERN (LPCC). MRM is supported by the UK Science and Technology Facilities Council (grant numbers ST/T001011/1 and ST/X000745/1).

\bibliography{references}

@article{Budnev:1975poe,
    author  = "Budnev, V. M. and Ginzburg, I. F. and Meledin, G. V. and Serbo, V. G.",
    title   = "{The Two photon particle production mechanism. Physical problems. Applications. Equivalent photon approximation}",
    journal = "Phys. Rept.",
    volume  = "15",
    pages   = "181--281",
    year    = "1975",
    doi     = "10.1016/0370-1573(75)90009-5"
}

@article{Drees:1988pp,
    author  = "Drees, Manuel and Zeppenfeld, Dieter",
    title   = "{Production of Supersymmetric Particles in Elastic $ep$ Collisions}",
    journal = "Phys. Rev. D",
    volume  = "39",
    pages   = "2536--2546",
    year    = "1989",
    doi     = "10.1103/PhysRevD.39.2536"
}

@article{Manohar:2016nzj,
    author        = "Manohar, Aneesh V. and Nason, Paolo and Salam, Gavin P. and Zanderighi, Giulia",
    title         = "{How bright is the proton? A precise determination of the photon parton distribution function}",
    eprint        = "1607.04266",
    archivePrefix = "arXiv",
    primaryClass  = "hep-ph",
    journal       = "Phys. Rev. Lett.",
    volume        = "117",
    number        = "24",
    pages         = "242002",
    year          = "2016",
    doi           = "10.1103/PhysRevLett.117.242002"
}

@article{Manohar:2017eqh,
    author        = "Manohar, Aneesh V. and Nason, Paolo and Salam, Gavin P. and Zanderighi, Giulia",
    title         = "{The photon content of the proton}",
    eprint        = "1708.01256",
    archivePrefix = "arXiv",
    primaryClass  = "hep-ph",
    journal       = "JHEP",
    volume        = "12",
    pages         = "046",
    year          = "2017",
    doi           = "10.1007/JHEP12(2017)046"
}

@article{Harland-Lang:2016apc,
    author        = "Harland-Lang, L. A. and Khoze, V. A. and Ryskin, M. G.",
    title         = "{The photon PDF in events with rapidity gaps}",
    eprint        = "1601.03772",
    archivePrefix = "arXiv",
    primaryClass  = "hep-ph",
    journal       = "Eur. Phys. J. C",
    volume        = "76",
    number        = "5",
    pages         = "255",
    year          = "2016",
    doi           = "10.1140/epjc/s10052-016-4100-2"
}

@article{Harland-Lang:2020veo,
    author        = "Harland-Lang, L. A. and Tasevsky, M. and Khoze, V. A. and Ryskin, M. G.",
    title         = "{A new approach to modelling elastic and inelastic photon-initiated production at the LHC: SuperChic 4}",
    eprint        = "2007.12704",
    archivePrefix = "arXiv",
    primaryClass  = "hep-ph",
    journal       = "Eur. Phys. J. C",
    volume        = "80",
    number        = "10",
    pages         = "925",
    year          = "2020",
    doi           = "10.1140/epjc/s10052-020-08455-0"
}

@article{Aad:2015bwa,
    author        = "Aad, Georges and others",
    collaboration = "ATLAS",
    title         = "{Measurement of exclusive $\gamma\gamma \to \ell^+\ell^-$ production in proton-proton collisions at $\sqrt{s}=7$ TeV with the ATLAS detector}",
    eprint        = "1506.07098",
    archivePrefix = "arXiv",
    primaryClass  = "hep-ex",
    reportNumber  = "CERN-PH-EP-2015-134",
    journal       = "Phys. Lett. B",
    volume        = "749",
    pages         = "242--261",
    year          = "2015",
    doi           = "10.1016/j.physletb.2015.07.069"
}

@article{CMS-TOTEM:2018hki,
    author        = "Sirunyan, Albert M. and others",
    collaboration = "CMS and TOTEM",
    title         = "{Observation of proton-tagged, central (semi)exclusive production of high-mass lepton pairs in pp collisions at 13 TeV with the CMS-TOTEM precision proton spectrometer}",
    eprint        = "1803.04496",
    archivePrefix = "arXiv",
    primaryClass  = "hep-ex",
    reportNumber  = "CMS-PPS-17-001, TOTEM-2018-001, CERN-EP-2018-014",
    journal       = "JHEP",
    volume        = "07",
    pages         = "153",
    year          = "2018",
    doi           = "10.1007/JHEP07(2018)153"
}

@article{Chatrchyan:2011ci,
    author        = "Chatrchyan, Serguei and others",
    collaboration = "CMS",
    title         = "{Exclusive $\gamma\gamma \to \mu^+\mu^-$ production in proton-proton collisions at $\sqrt{s}=7$ TeV}",
    eprint        = "1111.5536",
    archivePrefix = "arXiv",
    primaryClass  = "hep-ex",
    reportNumber  = "CERN-PH-EP-2011-187, CMS-FWD-10-005",
    journal       = "JHEP",
    volume        = "01",
    pages         = "052",
    year          = "2012",
    doi           = "10.1007/JHEP01(2012)052"
}

@article{Chatrchyan:2013foa,
    author        = "Chatrchyan, Serguei and others",
    collaboration = "CMS",
    title         = "{Study of exclusive two-photon production of $W^+W^-$ in $pp$ collisions at $\sqrt{s}=7$ TeV and constraints on anomalous quartic gauge couplings}",
    eprint        = "1305.5596",
    archivePrefix = "arXiv",
    primaryClass  = "hep-ex",
    reportNumber  = "CERN-PH-EP-2013-084, CMS-FSQ-12-010",
    journal       = "JHEP",
    volume        = "07",
    pages         = "116",
    year          = "2013",
    doi           = "10.1007/JHEP07(2013)116"
}

@article{CMS:2016exclWW,
    author        = "Khachatryan, Vardan and others",
    collaboration = "CMS",
    title         = "{Evidence for exclusive $\gamma\gamma \to W^+W^-$ production and constraints on anomalous quartic gauge couplings in $pp$ collisions at $\sqrt{s}=7$ and 8 TeV}",
    eprint        = "1604.04464",
    archivePrefix = "arXiv",
    primaryClass  = "hep-ex",
    reportNumber  = "CERN-EP-2016-073, CMS-FSQ-13-008",
    journal       = "JHEP",
    volume        = "08",
    pages         = "119",
    year          = "2016",
    doi           = "10.1007/JHEP08(2016)119"
}

@article{ATLAS:2020photonWW,
    author        = "Aad, Georges and others",
    collaboration = "ATLAS",
    title         = "{Observation of photon-induced $W^+W^-$ production in $pp$ collisions at $\sqrt{s}=13$ TeV using the ATLAS detector}",
    eprint        = "2010.04019",
    archivePrefix = "arXiv",
    primaryClass  = "hep-ex",
    reportNumber  = "CERN-EP-2020-165",
    journal       = "Phys. Lett. B",
    volume        = "816",
    pages         = "136190",
    year          = "2021",
    doi           = "10.1016/j.physletb.2021.136190"
}

@article{ATLAS:2020wny,
    author        = "Aad, Georges and others",
    collaboration = "ATLAS",
    title         = "{Observation and Measurement of Forward Proton Scattering in Association with Lepton Pairs Produced via the Photon Fusion Mechanism at ATLAS}",
    eprint        = "2009.14537",
    archivePrefix = "arXiv",
    primaryClass  = "hep-ex",
    reportNumber  = "CERN-EP-2020-173",
    journal       = "Phys. Rev. Lett.",
    volume        = "125",
    number        = "26",
    pages         = "261801",
    year          = "2020",
    doi           = "10.1103/PhysRevLett.125.261801"
}

@article{Vermaseren:1982cz,
    author  = "Vermaseren, J. A. M.",
    title   = "{Two photon processes at very high-energies}",
    journal = "Nucl. Phys. B",
    volume  = "229",
    pages   = "347--371",
    year    = "1983",
    doi     = "10.1016/0550-3213(83)90336-X"
}

@inproceedings{Baranov:1991,
    author    = "Baranov, S. P. and Duenger, O. and Shooshtari, H. and Vermaseren, J. A. M.",
    title     = "{LPAIR: A generator for lepton pair production}",
    booktitle = "{Proceedings of Physics at HERA}",
    volume    = "3",
    pages     = "1478",
    year      = "1991"
}

@article{Forthomme:2018sxa,
    author        = "Forthomme, Laurent",
    title         = "{CepGen -- A generic central exclusive processes event generator for hadron-hadron collisions}",
    eprint        = "1808.06059",
    archivePrefix = "arXiv",
    primaryClass  = "hep-ph",
    journal       = "Comput. Phys. Commun.",
    volume        = "271",
    pages         = "108225",
    year          = "2022",
    doi           = "10.1016/j.cpc.2021.108225"
}

@article{Archibald:2008zza,
    author  = "Archibald, J. and Gleisberg, T. and Hoche, S. and Krauss, F. and Schonherr, M. and Schumann, S. and Siegert, F. and Winter, J. C.",
    title   = "{Simulation of photon-photon interactions in hadron collisions with SHERPA}",
    journal = "Nucl. Phys. B Proc. Suppl.",
    volume  = "179-180",
    pages   = "218--225",
    year    = "2008",
    doi     = "10.1016/j.nuclphysbps.2008.07.027"
}

@article{Alwall:2014hca,
    author        = "Alwall, J. and Frederix, R. and Frixione, S. and Hirschi, V. and Maltoni, F. and Mattelaer, O. and Shao, H. S. and Stelzer, T. and Torrielli, P. and Zaro, M.",
    title         = "{The automated computation of tree-level and next-to-leading order differential cross sections, and their matching to parton shower simulations}",
    eprint        = "1405.0301",
    archivePrefix = "arXiv",
    primaryClass  = "hep-ph",
    journal       = "JHEP",
    volume        = "07",
    pages         = "079",
    year          = "2014",
    doi           = "10.1007/JHEP07(2014)079"
}

@article{Bellm:2015jjp,
    author        = "Bellm, Johannes and others",
    title         = "{Herwig 7.0/Herwig++ 3.0 release note}",
    eprint        = "1512.01178",
    archivePrefix = "arXiv",
    primaryClass  = "hep-ph",
    journal       = "Eur. Phys. J. C",
    volume        = "76",
    number        = "4",
    pages         = "196",
    year          = "2016",
    doi           = "10.1140/epjc/s10052-016-4018-8"
}

@article{Bewick:2023tfi,
    author        = "Bewick, Gavin and Ferrario Ravasio, Silvia and Gieseke, Stefan and Kiebacher, Stefan and Masouminia, Mohammad R. and Papaefstathiou, Andreas and Plätzer, Simon and Richardson, Peter and Samitz, Daniel and Seymour, Michael H. and Siódmok, Andrzej and Whitehead, James",
    title         = "{Herwig 7.3 release note}",
    eprint        = "2312.05175",
    archivePrefix = "arXiv",
    primaryClass  = "hep-ph",
    journal       = "Eur. Phys. J. C",
    volume        = "84",
    number        = "10",
    pages         = "1053",
    year          = "2024",
    doi           = "10.1140/epjc/s10052-024-13211-9"
}

@article{Bellm:2025pcw,
    author = "Bellm, J. and others",
    title = "{The Physics of Herwig 7}",
    eprint = "2512.16645",
    archivePrefix = "arXiv",
    primaryClass = "hep-ph",
    reportNumber = "CERN-TH-2025-252, IPPP-25-57, HERWIG-2025-01, KA-TP-36-2025, MCNET-25-31",
    month = "12",
    year = "2025"
}

@article{Bellm:2019zci,
    author = "Bellm, Johannes and others",
    title = "{Herwig 7.2 release note}",
    eprint = "1912.06509",
    archivePrefix = "arXiv",
    primaryClass = "hep-ph",
    reportNumber = "MAN/HEP/2019/011, CERN-TH-2019-213, IFJPAN-IV-2019-18, HERWIG-2019-02, UWTHPH-19-36, KA-TP-24-2019, LU-TP 19-57, MCnet-19-28, IPPP/19/91",
    doi = "10.1140/epjc/s10052-020-8011-x",
    journal = "Eur. Phys. J. C",
    volume = "80",
    number = "5",
    pages = "452",
    year = "2020"
}

@article{Sjostrand:2014zea,
    author        = "Sj{\"o}strand, Torbj{\"o}rn and Ask, Stefan and Christiansen, Jesper R. and Corke, Richard and Desai, Nishita and Ilten, Philip and Mrenna, Stephen and Prestel, Stefan and Rasmussen, Christine O. and Skands, Peter Z.",
    title         = "{An introduction to PYTHIA 8.2}",
    eprint        = "1410.3012",
    archivePrefix = "arXiv",
    primaryClass  = "hep-ph",
    journal       = "Comput. Phys. Commun.",
    volume        = "191",
    pages         = "159--177",
    year          = "2015",
    doi           = "10.1016/j.cpc.2015.01.024"
}

@article{Bierlich:2022pfr,
    author        = "Bierlich, Christian and others",
    title         = "{A comprehensive guide to the physics and usage of PYTHIA 8.3}",
    eprint        = "2203.11601",
    archivePrefix = "arXiv",
    primaryClass  = "hep-ph",
    journal       = "SciPost Phys. Codebases",
    pages         = "8",
    year          = "2022",
    doi           = "10.21468/SciPostPhysCodeb.8"
}

@article{Helenius:2018uul,
    author  = "Helenius, Ilkka",
    title   = "{Photon-photon and photon-hadron processes in Pythia 8}",
    journal = "CERN Proc.",
    volume  = "1",
    pages   = "119",
    year    = "2018",
    doi     = "10.23727/CERN-Proceedings-2018-001.119"
}

@article{Gleisberg:2008ta,
    author        = "Gleisberg, T. and H{\"o}che, Stefan and Krauss, F. and Sch{\"o}nherr, M. and Schumann, S. and Siegert, F. and Winter, J.",
    title         = "{Event generation with SHERPA 1.1}",
    eprint        = "0811.4622",
    archivePrefix = "arXiv",
    primaryClass  = "hep-ph",
    journal       = "JHEP",
    volume        = "02",
    pages         = "007",
    year          = "2009",
    doi           = "10.1088/1126-6708/2009/02/007"
}

@article{Bothmann:2019yzt,
    author        = "Bothmann, Enrico and others",
    collaboration = "Sherpa",
    title         = "{Event Generation with Sherpa 2.2}",
    eprint        = "1905.09127",
    archivePrefix = "arXiv",
    primaryClass  = "hep-ph",
    journal       = "SciPost Phys.",
    volume        = "7",
    number        = "3",
    pages         = "034",
    year          = "2019",
    doi           = "10.21468/SciPostPhys.7.3.034"
}

@article{Bothmann:2024yqu,
    author        = "Bothmann, Enrico and others",
    collaboration = "Sherpa",
    title         = "{Event generation with Sherpa 3}",
    eprint        = "2410.22148",
    archivePrefix = "arXiv",
    primaryClass  = "hep-ph",
    journal       = "JHEP",
    volume        = "12",
    pages         = "156",
    year          = "2024",
    doi           = "10.1007/JHEP12(2024)156"
}

@article{Hoche:2023mkv,
    author        = "H{\"o}che, Stefan and Krauss, Frank and Meinzinger, Peter",
    title         = "{Resolved photons in Sherpa}",
    eprint        = "2310.18674",
    archivePrefix = "arXiv",
    primaryClass  = "hep-ph",
    journal       = "Eur. Phys. J. C",
    volume        = "84",
    number        = "2",
    pages         = "178",
    year          = "2024",
    doi           = "10.1140/epjc/s10052-024-12551-w"
}

@article{Harland-Lang:2015cta,
    author        = "Harland-Lang, L. A. and Khoze, V. A. and Ryskin, M. G.",
    title         = "{Exclusive physics at the LHC with SuperChic 2}",
    eprint        = "1508.02718",
    archivePrefix = "arXiv",
    primaryClass  = "hep-ph",
    journal       = "Eur. Phys. J. C",
    volume        = "76",
    number        = "1",
    pages         = "9",
    year          = "2016",
    doi           = "10.1140/epjc/s10052-015-3832-8"
}

@article{Bahr:2008pv,
    author = "Bahr, M. and others",
    title = "{Herwig++ Physics and Manual}",
    eprint = "0803.0883",
    archivePrefix = "arXiv",
    primaryClass = "hep-ph",
    reportNumber = "CERN-PH-TH-2008-038, CAVENDISH-HEP-08-03, KA-TP-05-2008, DCPT-08-22, IPPP-08-11, CP3-08-05",
    doi = "10.1140/epjc/s10052-008-0798-9",
    journal = "Eur. Phys. J. C",
    volume = "58",
    pages = "639--707",
    year = "2008"
}

@article{Cridge:2021pxm,
    author        = "Cridge, T. and Harland-Lang, L. A. and Martin, A. D. and Thorne, R. S.",
    title         = "{QED parton distribution functions in the MSHT20 fit}",
    eprint        = "2111.05357",
    archivePrefix = "arXiv",
    primaryClass  = "hep-ph",
    journal       = "Eur. Phys. J. C",
    volume        = "82",
    number        = "1",
    pages         = "90",
    year          = "2022",
    doi           = "10.1140/epjc/s10052-022-10028-2"
}

@article{Masouminia:2021kne,
    author = "Masouminia, M. R. and Richardson, P.",
    title = "{Implementation of angularly ordered electroweak parton shower in Herwig 7}",
    eprint = "2108.10817",
    archivePrefix = "arXiv",
    primaryClass = "hep-ph",
    reportNumber = "IPPP/21/21",
    doi = "10.1007/JHEP04(2022)112",
    journal = "JHEP",
    volume = "04",
    pages = "112",
    year = "2022"
}

@article{Lee:2023hef,
    author = "Lee, Joon-Bin and Masouminia, M. R. and Seymour, Michael H. and Yang, Un-ki",
    title = "{Generalized angular-order parton showers in Herwig 7}",
    eprint = "2312.13125",
    archivePrefix = "arXiv",
    primaryClass = "hep-ph",
    reportNumber = "IPPP/23/73, MCnet-23-21",
    doi = "10.1007/JHEP08(2024)064",
    journal = "JHEP",
    volume = "08",
    pages = "064",
    year = "2024"
}

@article{Masouminia:2026udz,
    author = "Masouminia, M. R. and Richardson, P.",
    title = "{Gauge-invariant longitudinal modes in the Herwig 7 electroweak parton shower}",
    eprint = "2602.19927",
    archivePrefix = "arXiv",
    primaryClass = "hep-ph",
    reportNumber = "IPPP/25/91",
    doi = "10.1007/JHEP08(2026)160",
    journal = "JHEP",
    volume = "08",
    pages = "160",
    year = "2026"
}

@article{Kim:2026mfk,
    author = "Kim, Taehee and Lee, Joon-Bin and Masouminia, M. R. and Seymour, Michael H. and Yang, Un-ki",
    title = "{Searching for New Physics Inside Jets with the Herwig 7 Generalised Parton Shower}",
    eprint = "2604.00617",
    archivePrefix = "arXiv",
    primaryClass = "hep-ph",
    reportNumber = "CERN-TH-2026-074, MCnet-26-05, IPPP-26-15",
    month = "4",
    year = "2026"
}

@article{Darvishi:2021het,
    author = "Darvishi, Neda and Masouminia, M. R.",
    title = "{Electroweak radiative corrections in precision LHC measurements of W{\ensuremath{\pm}}/Z0+jets}",
    eprint = "2112.15487",
    archivePrefix = "arXiv",
    primaryClass = "hep-ph",
    reportNumber = "IPPP/21/59",
    doi = "10.1016/j.nuclphysb.2022.116025",
    journal = "Nucl. Phys. B",
    volume = "985",
    pages = "116025",
    year = "2022"
}

@article{Bierlich:2024Rivet,
    author = "Bierlich, Christian and Buckley, Andy and Butterworth, Jonathan and Gutschow, Christian and Lonnblad, Leif and Procter, Tomasz and Richardson, Peter and Yeh, Yoran",
    title = "{Robust Independent Validation of Experiment and Theory: Rivet version 4 release note}",
    eprint = "2404.15984",
    archivePrefix = "arXiv",
    primaryClass = "hep-ph",
    journal = "SciPost Phys. Codebases",
    volume = "36",
    year = "2024",
    doi = "10.21468/SciPostPhysCodeb.36"
}

@article{Buckley:2025YODA,
    author = "Buckley, Andy and Corpe, Louie and Filipovich, Matthew and Gutschow, Christian and Rozinsky, Nick and Thor, Simon and Yeh, Yoran and Yellen, Jamie",
    title = "{Consistent, multidimensional differential histogramming and summary statistics with YODA 2}",
    eprint = "2312.15070",
    archivePrefix = "arXiv",
    primaryClass = "hep-ph",
    journal = "SciPost Phys. Codebases",
    volume = "45",
    year = "2025",
    doi = "10.21468/SciPostPhysCodeb.45"
}

\end{document}